\documentclass [12pt]{article}
\usepackage{amssymb,amsfonts,latexsym,amsmath,amsthm,times}
\usepackage{graphicx}
\usepackage{epsfig}
\usepackage{fancyhdr}
\usepackage{color}
\usepackage[english]{babel}

\usepackage{bm}

\numberwithin{equation}{section}

\newcommand\beq{\begin{equation}}
\newcommand\eeq{\end{equation}}
\newcommand\beqa{\begin{eqnarray}}
\newcommand\eeqa{\end{eqnarray}}
\newcommand{\dd}{\text{d}}

\newcommand{\nn}{\nonumber\\}

\newcommand{\al}{\alpha}
\newcommand{\at}{\widetilde{a}}
\newcommand{\et}{\widetilde{\epsilon}}

\newcommand{\nutz}{\zeta}
\newcommand{\nuzo}{\nu_{0|1}}
\newcommand{\nuzt}{\nu_{0|2}}
\newcommand{\nuto}{\nu_{2|1}}
\newcommand{\nuzth}{\nu_{0|3}}
\newcommand{\nufz}{\nu_{4|0}}

\newcommand{\omegazt}{\omega_{0|2}}
\newcommand{\omeganu}{\nu}
\newcommand{\NS}{\text{NS}}

\newcommand{\zt}{\widetilde{\zeta}}

\newcommand{\nuhs}{\nu_{0}}
\newcommand{\one}{{(1)}}
\newcommand{\zero}{{(0)}}

\begin{document}

\footnotesize {\flushleft \mbox{\bf \textit{Math. Model. Nat.
Phenom.}}}
 \\
\mbox{\textit{{\bf Vol. 6, No. 4, 2011, pp. 37-76}}}

\thispagestyle{plain}

\vspace*{2cm} \normalsize \centerline{\Large \bf Hydrodynamics of inelastic Maxwell models}

\vspace*{1cm}

\centerline{\bf Vicente Garz\'o\footnote{E-mail: vicenteg@unex.es}
 and Andr\'es Santos\footnote{Corresponding
author. E-mail: andres@unex.es}}

\vspace*{0.5cm}

\centerline{Departamento de F\'{\i}sica, Universidad de
Extremadura, E-06071 Badajoz, Spain}


\vspace*{1cm}

\noindent {\bf Abstract.}
An overview of recent results pertaining to the hydrodynamic description (both Newtonian and non-Newtonian) of granular gases described by the  Boltzmann equation for inelastic Maxwell models is presented.
The use of this mathematical model allows us to get exact results for different problems. First, the Navier--Stokes constitutive equations with explicit expressions for the corresponding transport coefficients are derived by applying  the Chapman--Enskog method to inelastic gases. Second, the non-Newtonian rheological properties in the uniform shear flow (USF) are obtained in the steady state as well as in the transient unsteady regime. Next, an exact solution  for a special class of Couette flows characterized by a uniform heat flux is worked out. This solution shares the same rheological properties as the USF and, additionally, two generalized transport coefficients associated with the heat flux vector can be identified. Finally, the problem of small spatial perturbations of the USF is analyzed with a Chapman--Enskog-like method and generalized (tensorial) transport coefficients are obtained.

\vspace*{0.5cm}




\vspace*{1cm}

\setcounter{equation}{0}
\section{Introduction}

A simple and realistic physical model of a granular system under conditions of rapid flow consists of a fluid made  of inelastic hard spheres (IHS).
In the simplest version, the spheres are assumed to be smooth (i.e., frictionless) and the inelasticity in collisions  is accounted for by
 a constant coefficient of normal restitution $\alpha \leq 1$
\cite{BP04}.
At a kinetic theory level, all the relevant information about the dynamical properties of the fluid is embedded in the one-particle velocity distribution function $f(\mathbf{r}, \mathbf{v};t)$.
In the case of dilute gases, the conventional Boltzmann equation can be extended to the IHS model by changing the collision rules to account for the inelastic character of collisions \cite{BDS97,GS95}. On the other hand, even for ordinary gases made of elastic hard spheres ($\alpha=1$), the mathematical complexity of  the Boltzmann collision operator prevents one from obtaining exact results. These difficulties increase considerably in the  IHS case ($\alpha<1$). For instance, the fourth cumulant $a_2$ of the velocity distribution in the so-called homogeneous
cooling state  is not exactly known, although good estimates of it have been proposed
\cite{BP06,MS00,SM09,NE98}. Moreover, the explicit expressions for the Navier--Stokes (NS) transport
coefficients are not exactly known, but they have been approximately obtained by considering the leading terms in a Sonine polynomial expansion
\cite{BC01,BDKS98,GD99,GD02,GDH07,GHD07,GSM07,GVM09,Lutsko}.

As Maxwell already realized  in the context of elastic collisions \cite{Maxwell}, scattering laws where the collision rate of two particles
is independent of their relative velocity represent tractable mathematical models. In that case, the  change of velocity moments  of order $k$ per unit time can be expressed in terms of  moments of order $k'\leq k$,  without the explicit knowledge of the one-particle velocity
distribution function. In the conventional case of ordinary gases of particles colliding elastically, Maxwell
models correspond to particles interacting via a repulsive potential proportional to the inverse fourth power of distance (in three dimensions). However, in the framework of the Boltzmann equation one can introduce Maxwell models at the level of the cross section, without any reference to a specific interaction potential \cite{E81}.  Thanks to the use of Maxwell molecules, it is possible in some cases to find non-trivial exact solutions to the Boltzmann equation in far from equilibrium
situations \cite{E81,GS03,S09,SG95,TM80}.

Needless to say, the introduction of inelasticity through a constant coefficient of
normal restitution $\alpha\leq 1$, while keeping the independence of the collision rate with the relative velocity,  opens up new perspectives for exact results, including the elastic case ($\alpha=1$) as a
special limit. This justifies the growing interest in the so-called inelastic Maxwell models (IMM) by physicists and mathematicians alike in
the past few years
\cite{BMP02,BTE07,NK00,NK02a,NK02,NK03,BCG00,BC02,BC03,BCT03,BG06,BC07,BGM10,BE04,CCG00,C01,CDT05,EB02a,EB02,EB02bis,ETB06a,ETB06,G03,G07,GA05,GS07,KN02bis,MP02,MP02bis,S03,S07,SE03,SG07,SGV10,TK03}.
Furthermore, it is interesting to remark that recent
experiments \cite{KSSAON05} for magnetic grains with dipolar interactions are well described by IMM.
Apart from that, this mathematical model of granular gases allows one to explore the influence of inelasticity on the dynamic
properties in a clean way, without the need of introducing additional, and sometimes uncontrolled,
approximations.
Most of the studies devoted to IMM refer to homogeneous and isotropic states. In particular,  the high-velocity tails
\cite{BMP02,NK02a,EB02a,EB02,EB02bis,KN02bis} and the velocity cumulants \cite{BMP02,EB02,GA05,MP02,MP02bis,S03} have been derived. Nevertheless, much less is known about the hydrodynamic properties for inhomogeneous situations.

The aim of this review paper is to offer a brief survey on hydrodynamic properties recently derived in the context of IMM and also to derive some new results. Traditionally, hydrodynamics is understood as restricted to physical situations where the strengths of the spatial gradients of the hydrodynamic fields are small. This corresponds to the familiar NS constitutive equations for the momentum and heat fluxes. Notwithstanding this, it must be borne in mind that in granular gases there exists an inherent coupling between collisional dissipation and gradients, especially in steady states \cite{SGD04}. As a consequence, the NS description might fail  at finite inelasticity. This does not necessarily imply a failure of hydrodynamics in the sense that the state of the system is still characterized by the hydrodynamic fields but with constitutive relations more complex than the NS ones (non-Newtonian states). We address in this paper both views of hydrodynamics by  presenting the NS transport coefficients of IMM as well as some specific examples of non-Newtonian behavior.

The organization of this paper is as follows. The Boltzmann equation for IMM and the mass, momentum, and energy balance equations are presented in section \ref{sec2} Section \ref{sec3} deals with the first few collisional moments. In  sections \ref{sec4}--\ref{sec6} we  review the main results referring to the hydrodynamic properties of the IMM as a mathematical model of a granular gas. First, in section \ref{sec4} the Chapman--Enskog method is applied to the Boltzmann equation  and the NS transport coefficients are explicitly obtained without any approximation (such as truncation in Sonine polynomial expansions). Then, in sections \ref{sec5} and \ref{sec6} some  shear-flow  states where the NS description fails are analyzed and their non-Newtonian properties are exactly obtained.
In section \ref{sec7} we consider the generalized transport coefficients describing small spatial perturbations about the uniform shear flow. Finally, in section \ref{sec8} the main results are summarized and put in perspective.

\vspace*{0.5cm}
\setcounter{equation}{0}

\section{Inelastic Maxwell models}
\label{sec2}
The Boltzmann equation for IMM \cite{NK00,BCG00,CCG00,EB02} can be obtained from the Boltzmann equation
for IHS by replacing the term
$|\mathbf{g}\cdot \widehat{\bm{\sigma}}|$ in the collision rate
(where ${\bf g}={\bf v}_1-{\bf v}_2$ is the relative velocity of the
colliding pair and $\widehat{\bm{\sigma}}$ is the unit vector directed along
the centres of the two colliding spheres) by an {\em average\/} value proportional to the thermal velocity $\sqrt{2T/m}$ (where $T$ is
the granular temperature and $m$ is the mass of a particle). In the absence of external forces, the resulting
Boltzmann equation is \cite{EB02}
\beq
(\partial_t+{\bf v}\cdot\nabla)f({\bf r},{\bf
v};t)=J[{\bf r},{\bf v};t|f,f],
\label{1a}
\eeq
where
\beq
J[{\bf r},{\bf v}_1;t|f_1,f_2]=
\frac{\omeganu({\bf r},t)}{n({\bf r},t)\Omega_d}\int
\dd\widehat{\bm{\sigma}}\int \dd{\bf
v}_2
\left(\alpha^{-1}\widehat{b}^{-1}-1\right)
 f_1({\bf r},{\bf v}_1;t)f_2({\bf r},{\bf
v}_2;t)
\label{1b}
\eeq
is the IMM Boltzmann collision operator. Here,
\beq
n(\mathbf{r},t)=\int \dd\mathbf{v}\,
f({\bf r},{\bf v};t)
\label{b1}
\eeq
 is the number density, $\omeganu$ is an
effective collision frequency, $\Omega_d\equiv 2\pi^{d/2}/\Gamma(d/2)$ is the
total solid angle in $d$ dimensions, and $\alpha<1$ is the coefficient of normal
restitution. In addition,  $\widehat{b}$ is the operator transforming pre-collision
velocities into post-collision ones:
\beq
\widehat{b}{\bf v}_{1,2}={\bf v}_{1,2}\mp\frac{1+\alpha}{2}({\bf
g}\cdot\widehat{\bm{\sigma}})\widehat{\bm{\sigma}}.
\label{4}
\eeq
In Eq.\ (\ref{1b})  the collision rate is assumed to be independent of the
relative orientation between the unit vectors $\widehat{\mathbf{g}}$
and $\widehat{\bm{\sigma}}$ \cite{EB02}. In an alternative version  \cite{BCG00,BC02,BC03,BCT03,BG06,BC07,CCG00,C01}, the
collision rate has the same dependence on the scalar product
$\widehat{\mathbf{g}}\cdot \widehat{\bm{\sigma}}$ as in the case of
hard spheres. For simplicity, henceforth we will consider the version of IMM described by Eqs.\ (\ref{1a}) and (\ref{1b}).

The collision frequency $\omeganu$ is a free
parameter of the model that can be chosen to optimize the agreement for a given property between the IMM and the IHS model. In any case, one must have $\omeganu\propto n T^{1/2}$ to mimic the mean collision frequency of IHS,
where
\beq
T(\mathbf{r},t)=\frac{m}{dn(\mathbf{r},t)}\int \dd\mathbf{v}\,
{V}^2f({\bf r},{\bf v};t)
\label{b3}
\eeq
defines the granular temperature and
$\mathbf{V}=\mathbf{v}-\mathbf{u}$ is the peculiar velocity,
\beq
\mathbf{u}(\mathbf{r},t)=\frac{1}{n(\mathbf{r},t)}\int \dd\mathbf{v}\,
\mathbf{v}f({\bf r},{\bf v};t)
\label{b4}
\eeq
being the flow velocity.

In  a hydrodynamic description of an ordinary gas the state of the system is defined by the  fields associated with  the local densities of mass, momentum, and energy, conventionally chosen as $n$, $\mathbf{u}$, and $T$. For granular gases, even though kinetic energy is not conserved  upon collisions, it is adequate to take these quantities as hydrodynamic fields \cite{D01,DB06}. The starting point is the set of macroscopic balance equations which follow directly from Eq.~(\ref{1a}) by multiplying it by $\{1,\mathbf{v},V^2\}$ and integrating over velocity. These balance equations read
\beq
D_t n+n\nabla\cdot \mathbf{u}=0,
\label{b7}
\eeq
\beq
D_t\mathbf{u}+\frac{1}{mn}\nabla\cdot\mathsf{P}=\mathbf{0},
\label{b8}
\eeq
\beq
D_tT+\frac{2}{dn}\left(\nabla\cdot\mathbf{q}+\mathsf{P}:\nabla
\mathbf{u}\right)=-\zeta T.
\label{b9}
\eeq
In these equations, $D_t\equiv\partial_t+\mathbf{u}\cdot\nabla$ is the
material time derivative,
\beq
\mathsf{P}(\mathbf{r},t)=m\int\dd\mathbf{v}\, \mathbf{V}\mathbf{V}f(\mathbf{r},\mathbf{v};t)
\label{b10}
\eeq
is the pressure tensor,
\beq
\mathbf{q}(\mathbf{r},t)=\frac{m}{2}\int\dd\mathbf{v}\,
V^2 \mathbf{V}f(\mathbf{r},\mathbf{v};t)
\label{b11}
\eeq
is the heat flux, and
\beq
\zeta(\mathbf{r},t)=
-\frac{m}{d n(\mathbf{r},t)T(\mathbf{r},t)}
\int \dd\mathbf{v}\, {V}^2J[{\bf r},{\bf v};t|f,f]
\label{b12}
\eeq
is the cooling rate.
The energy balance equation (\ref{b9}) differs from that of an ordinary gas by the presence of the sink term $-\zeta T$ measuring the rate of energy dissipation due to collisions.

The set of balance equations (\ref{b7})--(\ref{b9}) are
generally valid, regardless of the details of the inelastic model  and so their structure is common for both IMM and IHS.
It is apparent that Eqs.\ (\ref{b7})--(\ref{b9}), while exact, do not constitute a closed set of equations. To close them and get a \textit{hydrodynamic} description, one has to express the momentum and heat fluxes in terms of the hydrodynamic fields. These relations are called constitutive equations. In their more general form, the fluxes are expressed as \textit{functionals} of the hydrodynamic fields, namely
\beq
\mathsf{P}=\mathsf{P}[n,\mathbf{u},T],\quad \mathbf{q}=\mathbf{q}[n,\mathbf{u},T].
\label{hydro}
\eeq
In other words, all the space and time dependence of the pressure tensor and the heat flux occurs through a functional dependence on $n$, $\mathbf{u}$, and $T$, not necessarily local in space or time. However, for sufficiently small spatial gradients, the functional dependence  can be assumed to be local in time and weakly non-local in space. More specifically,
\beq
P_{ij}=p-\eta_\NS\left(\nabla_i u_j+\nabla_j u_i-\frac{2}{d}\nabla\cdot\mathbf{u}\delta_{ij}\right),
\label{Newton}
\eeq
\beq
\mathbf{q}=-\kappa_\NS \nabla T-\mu_\NS\nabla n.
\label{Fourier}
\eeq
In Newton's equation \eqref{Newton} and Fourier's equation \eqref{Fourier}, $p=nT=\frac{1}{d}\text{Tr }\mathsf{P}$ is the hydrostatic pressure,  $\eta_\NS$ is the shear viscosity, $\kappa_\NS$ is the thermal conductivity coefficient, and $\mu_\NS$ is a transport coefficient that vanishes in the elastic case. When the constitutive equations \eqref{Newton} and \eqref{Fourier} are introduced into the momentum and energy balance equations, the set  \eqref{b7}--(\ref{b9}) becomes closed and one arrives to the familiar NS hydrodynamic equations.

It must be noticed that, in the case of IHS, the cooling rate $\zeta$ also has to be expressed as a functional of the hydrodynamic fields. However, in the case of IMM, $\zeta$ is just proportional to the effective collision frequency $\omeganu$. More specifically \cite{GS07,S03},
\beq
\zeta=\frac{1-\alpha^2}{2d}\omeganu.
\label{10}
\eeq
This equation allows one to fix $\omeganu$ under the criterion that the cooling rate of IMM be the same as that of IHS of diameter $\sigma$. When the IHS cooling rate is evaluated in the Maxwellian approximation \cite{GS95,NE98}, one gets
\begin{equation}
\label{4ba}
\omeganu=\frac{d+2}{2}\nuhs,
\end{equation}
where
\beq
\label{4bb}
\nuhs=\frac{4\Omega_d}{\sqrt{\pi}(d+2)}n\sigma^{d-1}\sqrt{\frac{T}{m}}.
\end{equation}
The collision frequency $\nuhs$ is the one associated with the NS shear viscosity of an ordinary gas ($\alpha=1$) of both Maxwell molecules and hard spheres, i.e., $\eta_{\NS}=p/\nuhs\equiv \eta_0$ at $\alpha=1$. However, the specific form \eqref{4bb} will not be needed in the remainder of the paper.

An important problem in monocomponent systems is the self-diffusion  process. In that case one assumes that some particles are labeled with a tag but otherwise they are mechanically equivalent to the untagged particles. The balance equation reflecting the conservation of mass for the tagged  particles is
\beq
D_t x_1+\frac{1}{nm}\nabla\cdot \mathbf{j}_1=0,
\label{tag}
\eeq
where $x_1=n_1/n$ is the mole fraction of the tagged particles and
\beq
\mathbf{j}_1=m\int\dd\mathbf{V}\, \mathbf{V}f_1(\mathbf{V})
\label{j0}
\eeq
is the mass flux of the tagged particles. Analogously to the case of Eqs.\ \eqref{b7}--\eqref{b9}, one needs a constitutive equation for $\mathbf{j}_1$ to get a closed set of equations. For small spatial gradients, Fick's law applies, i.e.,
\beq
\mathbf{j}_1=-D_\NS\nabla x_1,
\label{Fick}
\eeq
where $D_\NS$ is the self-diffusion coefficient.

The Boltzmann equation for IMM, Eqs.\ \eqref{1a} and \eqref{1b}, refers to a monodisperse gas. The extension to a multi-component gas is straightforward \cite{NK02,G03,GA05}.  Instead of a single collision frequency $\nu$, one has a set of collision frequencies $\nu_{ij}$ that can be chosen to reproduce the cooling rates $\zeta_{ij}$ of IHS evaluated in a two-temperature Maxwellian approximation. In that case, the result is \cite{GD99b}
\beq
\label{4bc}
\omeganu_{ij}=\frac{\Omega_d}{\sqrt{\pi}}n_j\sigma_{ij}^{d-1}\sqrt{2\left(\frac{T_i}{m_i}+\frac{T_j}{m_j}\right)},
\end{equation}
where $\sigma_{ij}=(\sigma_{i}+\sigma_{j})/2$ and $T_i$ is the partial granular temperature of species $i$.

\vspace*{0.5cm}
\setcounter{equation}{0}

\section{Collisional moments}
\label{sec3}
As said in the Introduction, the key advantage of the Boltzmann equation for Maxwell
models (both elastic and inelastic) is that the (collisional)
moments of the operator $J[f,f]$ can be exactly evaluated in terms of the moments of
$f$, without the explicit knowledge of the latter \cite{TM80}.  More explicitly, the collisional moments of order $k$ are given by a bilinear combination of moments of order $k'$ and $k-k'$ with $0\leq k'\leq k$. In particular, the second- and third-order collisional moments are \cite{GS07}
\beq
m\int\dd \mathbf{V} \, V_iV_j J[\mathbf{V}|f,f]=-\nuzt(P_{ij}-p\delta_{ij})-\nutz p \delta_{ij},
\label{Z1}
\eeq
\beq
\frac{m}{2}\int\dd \mathbf{V} \, V_iV_jV_k J[\mathbf{V}|f,f]=-\nuzth Q_{ijk}-\frac{\nuto-\nuzth}{d+2}\left(q_i\delta_{jk}+q_j\delta_{ik}+q_k\delta_{ij}\right).
\label{Z2}
\eeq
In Eq.\ \eqref{Z1}, the cooling rate $\zeta$ is given by Eq.\ \eqref{10} and
\beq
\nuzt=\nutz+\frac{(1+\alpha)^2}{2(d+2)}\nu.
\label{X6}
\eeq
As for Eq.\ \eqref{Z2}, one has
\beq
 \nuto=\frac{3}{2}\nutz+\frac{(1+\al)^2(d-1)}{2d(d+2)}\nu,\quad \nuzth=\frac{3}{2}\nuzt.
\label{X6b}
\eeq
Moreover,
\beq
Q_{ijk}=\frac{m}{2}\int \dd\mathbf{V}\, V_iV_jV_k f(\mathbf{V})
\label{Z3}
\eeq
is a third-rank tensor, whose trace is the heat flux. In particular, from Eq.\ \eqref{Z2} we easily get
\beq
\frac{m}{2}\int\dd \mathbf{V} \, V^2V_i J[\mathbf{V}|f,f]=-\nuto q_i.
\label{Z2b}
\eeq

The evaluation of the fourth-order collisional moments is more involved and their expressions can be found in Ref.\ \cite{GS07}. Here, for the sake of illustration, we only quote the equation related to the isotropic moment \cite{BC07,GS07}:
\beq
m\int\dd \mathbf{V} \, V^4 J[\mathbf{V}|f,f]=-\nufz M_{4}+\frac{\lambda_1}{nm}d^2p^2-\frac{\lambda_2}{nm}\Pi_{ij}\Pi_{ji},
\label{X9}
\eeq
where
\beq
M_4=m \int\dd \mathbf{V} \, V^4 f(\mathbf{V})
\eeq
is the isotropic fourth-order moment and $\Pi_{ij}=P_{ij}-p\delta_{ij}$ is the irreversible part of the pressure tensor.  The coefficients in Eq.\ \eqref{X9} are
\beq
\nufz=2\nutz+\frac{(1+\al)^2\left(4d-7+6\alpha-3\alpha^2\right)}{8d(d+2)}\nu,
\label{X10}
\eeq
\beq
\lambda_1=\frac{(1+\al)^2\left(4d-1-6\al+3\al^2\right)}{8d^2}\nu,
\label{X14}
\eeq
\beq
\lambda_2=\frac{(1+\al)^2\left(1+6\al-3\al^2\right)}{4d(d+2)}\nu.
\label{X13}
\eeq
In Eqs.\ \eqref{X6}, \eqref{X6b}, and \eqref{X10} the collision frequencies $\nuzt$,  $\nuto$, and $\nufz$  have been decomposed into a part
inherent to the collisional cooling plus a genuine part associated with  the  collisional transfers.

In self-diffusion problems it is important to know the first-order collisional moment of $J[f_1,f_2]$ with $f_1\neq f_2$. After simple algebra one gets \cite{G03,GA05}
\beq
m\int\dd \mathbf{V} \,\mathbf{V} J[\mathbf{V}|f_1,f_2]=-\nuzo\left(\mathbf{j}_1-\mathbf{j}_2\right),
\label{Jji}
\eeq
where
\beq
\nuzo=\frac{1+\alpha}{2d}\nu
\label{nuzo}
\eeq
and $\mathbf{j}_s$ ($s=1,2$) is defined by Eq.\ \eqref{j0}.

Before studying the hydrodynamic properties of IMM, it is convenient to briefly analyze the so-called homogeneous cooling state (HCS). This is the simplest situation of a granular gas and, additionally, it plays the role of the reference state around which to carry out the Chapman--Enskog expansion.
The HCS is an isotropic, spatially uniform  free cooling state \cite{BP04}, so the
Boltzmann equation \eqref{1a} becomes
\beq
\partial_t f({v},t)=J[{v}|f,f],
\label{Z1x}
\eeq
which must be complemented with a  given initial condition $f({v},0)$. Since the
collisions are inelastic, the granular temperature $T(t)$ monotonically decays in time and so a steady state
does not exist. In fact, the  mass and momentum balance equations \eqref{b7} and \eqref{b8} are trivially satisfied and the energy equation \eqref{b9} reduces to
\beq
\partial_t T=-\zeta T,
\label{3.12}
\eeq
whose solution is given by Haff's law \cite{Haff}, namely
\beq
T(t)=\frac{T(0)}{\left[1+\frac{1}{2}\zeta(0)t\right]^2},
\label{Haff}
\eeq
where we have taken into account that $\zeta\propto nT^{1/2}$. The next non-trivial isotropic moment is $M_4(t)$. The evolution equation for the \textit{reduced}
moment
\beq
M_4^*(t)=\frac{M_4(t)}{nm[2T(t)/m]^2}
\eeq
is
\beq
\partial_t M_{4}^*=-\left(\nufz-2\nutz\right) M_{4}^*+\lambda_1
\frac{d^2}{4}.
\label{Z8}
\eeq
Note that, while Eqs.\ \eqref{3.12} and \eqref{Haff} are valid both for IHS and IMM, Eq.\ \eqref{Z8} is restricted to IMM. The general solution of Eq.\ \eqref{Z8} is
\beq
M_4^*(t)=\left[M_4^*(0)-\mu_4\right]\left[1+\frac{1}{2}\zeta(0)t\right]^{-4(\nufz/2\nutz-1)}+\mu_4,
\label{M4}
\eeq
where
\beq
\mu_4\equiv \frac{d^2}{4}\frac{\lambda_1}{\nufz-2\nutz}.
\eeq

\begin{figure}[htbp]
\centerline{\includegraphics[width= .5\columnwidth]{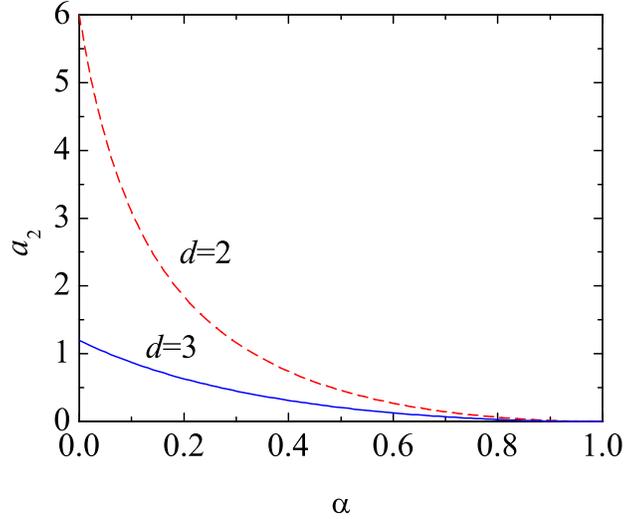}}
\caption{Plot of the fourth cumulant $a_2$ in the HCS for $d=2$ (dashed
line) and $d=3$ (solid line).\label{fig1}}
\end{figure}

In the one-dimensional case, Eq.\ \eqref{X10} shows that the difference $\nufz-2\nutz=-(1-\al^2)^2\omeganu/8$ is negative definite for any $\alpha<1$, so that, according to Eq.\ \eqref{M4}, the scaled moment $M_4^*$ diverges with time. On the other hand, if $d\geq 2$, $\nufz-2\nutz>0$ for any $\alpha$ and, consequently, the moment $M_4^*$ goes asymptotically to the value $\mu_4$. In that case, the corresponding fourth cumulant defined by
\beq
a_2\equiv\frac{4}{d(d+2)}M_4^*-1
\label{a2A}
\eeq
 is given by
\beq
a_2=\frac{6(1-\alpha)^2}{4d-7+3\alpha(2-\alpha)}.
\label{a2B}
\eeq
The dependence of $a_2$ on the coefficient of restitution $\alpha$ for $d=2$ and $3$ is displayed in Fig.\ \ref{fig1}. It is apparent that $a_2$ is always positive and rapidly grows with inelasticity, especially in the two-dimensional case. In contrast, the dependence of the IHS $a_2$ on $\alpha$ is non-monotonic and  much weaker \cite{SM09}.

It has been proven \cite{BC03,BCT03} that, provided that
$f({v},0)$ has a finite moment of some order higher than two, $f({v},t)$ asymptotically tends
toward a self-similar solution of the form
\beq
f({v},t)\to n \left[\frac{m}{2T(t)}\right]^{d/2} \phi_h(c(t)),\quad \mathbf{c}(t)\equiv \frac{\mathbf{v}}{\sqrt{2T(t)/m}},
\label{3.1}
\eeq
where  $\phi_h(c)$ is an  isotropic distribution. This scaled distribution
is only exactly known in the one-dimensional case \cite{BMP02}, where it is given by
\beq
\phi_h(c)=\frac{2^{3/2}}{\pi}\frac{1}{(1+2c^2)^2}.
\label{Z18}
\eeq
All the moments of this Lorentzian form of order higher than two are divergent. This is consistent with the divergence of $M_4^*(t)$ found in Eq.\ \eqref{M4}.
It is interesting to note that  if the initial
state is anisotropic then the  anisotropy does not vanish in the scaled velocity
distribution function for long times \cite{GS07}. As a consequence, while the distribution \eqref{Z18} represents
the asymptotic form $\phi_h(c)$ for a wide class of isotropic initial conditions, it cannot be reached,
strictly speaking, from any anisotropic initial state. Whether or not there exists a generalization
of \eqref{Z18} for anisotropic states is, to the best of our knowledge, an open problem.

Although the explicit expression of $\phi_h(c)$ is not known for $d\geq 2$, its high-velocity tail has been found to be of the form \cite{NK02a,EB02a,EB02,KN02bis}
\beq
\phi_h({c})\sim c^{-d-s(\alpha)},
\label{18}
\eeq
where the exponent $s(\alpha)$ is the solution of the transcendental
equation
\beq
1-\frac{1-\alpha^2}{4d}
s=_2\!\!F_1\left[-\frac{s}{2},\frac{1}{2};\frac{d}{2};\frac{3+2\alpha-\alpha^2}{4}\right]+\left(\frac{1+\alpha}{2}\right)^s\frac{\Gamma(\frac{s+1}{2})\Gamma(\frac{d}{2})}{\Gamma(\frac{s+d}{2})\Gamma(\frac{1}{2})},
\label{19}
\eeq
$_2F_1[a,b;c;z]$ being a hypergeometric function \cite{AS72}. Equation
\eqref{18} implies that those moments of order $k\geq
s(\alpha)$ are divergent.

The evolution of moments of order equal to or lower than four for \textit{anisotropic} initial states has been analyzed in Ref.\ \cite{GS07}.

\vspace*{0.5cm}
\setcounter{equation}{0}

\section{Navier--Stokes hydrodynamic description\label{sec4}}

The standard Chapman--Enskog method \cite{CC70} can be generalized to
inelastic collisions \cite{BP04} to obtain the dependence of the
NS transport coefficients on the coefficient of restitution from the Boltzmann equation
\cite{BC01,BDKS98,GD02,GM02,GSM07,GVM09} and from the Enskog equation \cite{GD99,GDH07,GHD07,Lutsko}. Here the
method will be applied to the Boltzmann equation (\ref{1a}) for IMM.

In order to get the hydrodynamic description in the sense of Eq.\ \eqref{hydro}, we need to obtain a \textit{normal} solution to the Boltzmann equation. A normal solution is a special solution where all the space and time dependence of the velocity distribution function takes place via a functional dependence on the hydrodynamic fields, i.e.,
\beq
f=f[\mathbf{v}|n,\mathbf{u},T].
\label{normal}
\eeq
This functional dependence can be made explicit by the Chapman--Enskog method if the gradients are small.
In the  method, a factor $\epsilon$ is assigned to every
gradient operator and  the distribution function is
represented as a series in this formal ``uniformity'' parameter,
\beq
f=f^\zero+\epsilon f^\one+\epsilon^2 f^{(2)}+\cdots.
\label{c1}
\eeq
Insertion of this expansion in the definitions of the fluxes (\ref{b10}) and
(\ref{b11})  gives the corresponding expansion
for these quantities. Finally, use of these expansions in the balance equations
(\ref{b7})--(\ref{b9}) leads to an identification of the time derivatives of
the fields as an expansion in the gradients,
\beq
\partial_t=\partial_t^\zero+\epsilon
\partial_t^\one+\epsilon^2\partial_t^{(2)}+\cdots.
\label{c2}
\eeq

The starting point is the zeroth order solution. The macroscopic balance equations to zeroth order are
\beq
\partial_t^\zero n=0,\quad \partial_t^\zero \mathbf{u}=\mathbf{0}, \quad
\partial_t^\zero T=-\zeta T.
\label{c3}
\eeq
Here, we have taken into account that in the Boltzmann operator (\ref{1b}) the
effective collision frequency $\omeganu\propto n T^{1/2}$, and hence the cooling rate $\zeta$  is  a
functional of $f$ only through the density $n$ and granular temperature $T$ [see Eq.\ \eqref{10}].
Consequently,
$\zeta^\zero=\zeta$.
To zeroth order in the gradients the kinetic equation (\ref{1a}) reads
\beq
\partial_t f^\zero=J[\mathbf{V}|f^\zero,f^\zero].
\label{c4}
\eeq
This equation coincides with that of the HCS, Eq.\ \eqref{Z1x}. Moreover, since $f^\zero$ must be a normal solution, its temporal dependence only occurs through temperature and so it is given by the right-hand side of Eq.\ \eqref{3.1}, except that $n\to n(\mathbf{r},t)$ and $T\to
T(\mathbf{r},t)$ are local quantities and $\mathbf{v}\to
\mathbf{V}=\mathbf{v}-\mathbf{u}(\mathbf{r},t)$. The normal character of $f^\zero$ allows one to write
\beq
\partial_t^\zero f^\zero(\mathbf{V})=-\zeta T
\frac{\partial}{\partial T}f^\zero(\mathbf{V})
=\frac{\zeta}{2}\frac{\partial}{\partial \mathbf{V}}\cdot
\mathbf{V}f^\zero(\mathbf{V}),
\label{c5}
\eeq
so that Eq.\ \eqref{c4} becomes
\beq
\frac{\zeta}{2}\frac{\partial}{\partial \mathbf{V}}\cdot
\mathbf{V}f^\zero(\mathbf{V})=J[\mathbf{V}|f^\zero,f^\zero].
\label{c4x}
\eeq

Since $f^\zero$ is isotropic, it follows that
\beq
{P}_{ij}^\zero=p\delta_{ij},\quad \mathbf{q}^\zero=\mathbf{0}.
\label{c6}
\eeq
Therefore, the macroscopic balance equations to first order give
\beq
D_t^\one n=-n\nabla\cdot\mathbf{u},\quad D_t^\one\mathbf{u}=-\frac{\nabla
p}{mn},\quad D_t^\one T=-\frac{2T}{d}\nabla\cdot\mathbf{u},
\label{c7}
\eeq
where $D_t^\one\equiv \partial_t^\one+\mathbf{u}\cdot \nabla$.
To first order in the gradients, Eq.~(\ref{1a}) leads to the following
equation for $f^\one$:
\beq
\left(\partial_t^\zero+\mathcal{L}\right)f^\one(\mathbf{V})=-\left(D_t^\one
+\mathbf{V}\cdot\nabla\right)f^\zero(\mathbf{V}),
\label{47}
\eeq
where $\mathcal{L}$ is the linearized collision operator
\beqa
\mathcal{L}f^\one(\mathbf{V}_1)&=&-\left(J[f^\zero,f^\one]+J[f^\one,f^\zero]\right)\nn
&=&
-\frac{\omeganu}{n\Omega_d}\int
\dd\widehat{\bm{\sigma}}\int \dd{\bf V}_2
\left(\alpha^{-1}\widehat{b}^{-1}-1\right)
\left[ f^\zero({\bf V}_1)f^\one({\bf V}_2)+f^\zero({\bf V}_2)f^\one({\bf
V}_1)\right].\nn
\label{c8}
\eeqa

Using (\ref{c7}), the right-hand side of Eq.~(\ref{47}) can evaluated explicitly, so the integral equation for $f^\one$ can be written as
\beq
\left(\partial_t^\zero+\mathcal{L}\right)f^\one(\mathbf{V})=
{\bf A}(\mathbf{V})\cdot \nabla
\ln T+{\bf B}(\mathbf{V})\cdot \nabla \ln n+
\mathsf{C}(\mathbf{V}):\nabla \mathbf{u},
\label{c12}
\eeq
where
\beq
{\bf A}\equiv\frac{\bf V}{2}\frac{\partial}{\partial {\bf V}}\cdot \left({\bf
V}f^\zero\right)-\frac{T}{m}\frac{\partial}{\partial {\bf V}}f^\zero,
\label{48}
\eeq
\beq
{\bf B}\equiv-{\bf V} f^\zero-\frac{T}{m}\frac{\partial}{\partial {\bf V}}f^\zero,
\label{49}
\eeq
\beq
C_{ij}\equiv\frac{\partial}{\partial V_i}\left(V_j
f^\zero\right)-\frac{1}{d}\delta_{ij} \frac{\partial}{\partial {\bf V}}\cdot
\left({\bf
V}f^\zero\right).
\label{50}
\eeq
The structure of Eq.\ \eqref{c12} is identical to that of IHS, except for the detailed form \eqref{c8} of the linearized Boltzmann collision operator $\mathcal{L}$.
In spite of the advantages of IMM, Eq.\ \eqref{c12} is mathematically rather intricate and its solution is not known. In the case of IHS, a trial function (based on a truncated Sonine polynomial expansion) for $f^\one$ is proposed. The coefficients in the trial function, which are directly related to the NS transport coefficients, are  obtained in an approximate way by taking velocity moments.
On the other hand, the use of a trial function is not needed in the case of IMM and the NS transport coefficients can be obtained exactly. The key point is that, upon linearization of Eqs.\ \eqref{Z1} and \eqref{Z2b}, one has
\beq
m\int \dd\mathbf{V}\,{V_i}{V_j}\mathcal{L}f^\one(\mathbf{V})=\nuzt
{P}_{ij}^\one,
\label{c9}
\eeq
\beq
\frac{m}{2}\int
\dd\mathbf{V}\,V^2\mathbf{V}\mathcal{L}f^\one(\mathbf{V})=\nuto \mathbf{q}^\one.
\label{c10}
\eeq

Now we multiply both sides of Eq.\ (\ref{c12}) by $m V_i V_j$ and integrate
over ${\bf V}$ to obtain
\beq
(\partial_t^\zero+\nuzt)P_{ij}^\one=-p\left(\nabla_i u_j+\nabla_j u_i-\frac{2}{d}\nabla\cdot\mathbf{u}\delta_{ij}\right).
\label{52}
\eeq
This equation shows that $P_{ij}^\one$ is proportional to the right-hand side divided by a collision frequency. Therefore, $P_{ij}^\one\propto p/nT^{1/2}=T^{1/2}$ and so
\beq
\partial_t^\zero \mathsf{P}^\one=-\frac{\zeta}{2}\mathsf{P}^\one,
\label{c16}
\eeq
where we have taken into account Eq.\ \eqref{c3}.
As a consequence, the solution to Eq.~(\ref{52}) is
\beq
P_{ij}^\one=-\eta_\NS \left(\nabla_i u_j+\nabla_j u_i-\frac{2}{d}\nabla\cdot\mathbf{u}\delta_{ij}\right),
\label{53}
\eeq
where
\beq
\eta_\NS=\frac{p}{\nuzt-\frac{1}{2}\zeta}.
\label{54}
\eeq
Comparison with Eq.\ \eqref{Newton} allows one to identify Eq.\ \eqref{54} with the NS shear viscosity of IMM.

Let us consider next the heat flux.
Multiplying both sides of Eq.\ (\ref{c12}) by $\frac{1}{2}m V^2\mathbf{V}$
and integrating over ${\bf V}$ we get
\beq
\left(\partial_t^\zero+\nuto
\right){\bf q}^\one=-\frac{d+2}{2}(1+2a_2)\frac{p}{m} \nabla T-\frac{d+2}{2}a_2\frac{T^2}{m}\nabla n ,
\label{60}
\eeq
where use has been made of Eq.\ \eqref{c10}. Here, $a_2$ is the fourth cumulant of $f^\zero$, whose expression is given by Eq.\ \eqref{a2B}.
The right-hand side of Eq.\ \eqref{60} implies that the heat flux has the structure
\beq
{\bf q}^\one=-\kappa_\NS \nabla T-\mu_\NS \nabla n.
\label{61}
\eeq
By dimensional analysis,
$\kappa_\NS\propto T^{1/2}$ and $\mu_\NS\propto T^{3/2}$. Consequently,
\beqa
\partial_t^\zero {\bf q}^\one &=&\frac{\zeta}{2} \kappa_\NS\nabla
T+\frac{3\zeta}{2} \mu_\NS \nabla n+ \kappa_\NS\nabla \zeta
T\nonumber\\
&=&\zeta\left[2 \kappa_\NS\nabla T+\left(\frac{3}{2}\mu_\NS+\kappa_\NS
\frac{T}{n}\right)\nabla n\right],
\label{62}
\eeqa
where in the last step we have taken into account that $\zeta\propto
nT^{1/2}$.
Inserting this equation into Eq.\ (\ref{60}), one can identify the transport
coefficients as
\beq
\kappa_\NS=\frac{p}{m}\frac{d+2}{2}\frac{1+2a_2}{\nuto-2\zeta},
\label{63}
\eeq
\beq
\mu_\NS=\frac{T}{n}\frac{\kappa_\NS}{1+2a_2}
\frac{\zeta+a_2
\nuto}{\nuto-\frac{3}{2}\zeta}.
\label{c19}
\eeq

Equations \eqref{54}, \eqref{63}, and \eqref{c19} provide the NS transport coefficients of the granular gas modeled by the IMM in terms of the cooling rate $\zeta$, the collision frequencies $\nuzt$ and $\nuto$, and the HCS fourth cumulant $a_2$. Making use of their explicit expressions, Eqs.\ \eqref{10}, \eqref{X6}, \eqref{X6b}, and \eqref{a2B}, respectively, the $\alpha$-dependence of the transport coefficients is given by
\beq
\eta_\NS=\eta_0\frac{8d}{(1+\alpha)\left[3d+2+(d-2)\alpha\right]},
\label{etaNS}
\eeq
\beq
\kappa_\NS=\kappa_0\frac{8(d-1)\left[5+4d-9\alpha(2-\alpha)\right]}{(1+\alpha)\left(d-4+3d \alpha\right)\left[4d-7+3\alpha(2-\alpha)\right]},
\label{kappaNS}
\eeq
\beq
\mu_\NS=\kappa_0\frac{T}{n}\frac{16(1-\alpha)\left[2d^2+8d-1-6(d+2)\alpha+9\alpha^2\right]}{(1+\alpha)^2\left(d-4+3d \alpha\right)\left[4d-7+3\alpha(2-\alpha)\right]},
\label{muNS}
\eeq
where $\eta_0=(d+2)(p/2\nu)=p/\nuhs$ and $\kappa_0=[d(d+2)/2(d-1)](\eta_0/m)$ are the NS shear viscosity and thermal conductivity coefficients in the elastic limit ($\alpha=1$), respectively.

It is interesting to rewrite Eq.\ \eqref{61} using  $T$ and $nT^{1/2}$  as hydrodynamic variables instead of $T$ and $n$ \cite{GSM07,MSG07}. In fact,  both the collision frequency $\nu$ and the cooling rate $\zeta$ are proportional to $nT^{1/2}$. In these variables, the heat flux becomes
\beq
{\bf q}^\one=-\kappa'_\NS \nabla T-\mu_\NS T^{-1/2}\nabla \left(nT^{1/2}\right),
\label{qone}
\eeq
where $\kappa'_\NS=\kappa_\NS-\mu_\NS (n/2T)$. Inserting Eqs.\ \eqref{63} and \eqref{c19}, we easily get
\beqa
\kappa'_\NS&=&\frac{p}{m}\frac{d+2}{2}\frac{1+\frac{3}{2}a_2}{\nuto-\frac{3}{2}\zeta}\nn
&=&\kappa_0\frac{8\left[1+2d-3\alpha(2-\alpha)\right]}{(1+\alpha)^2\left[4d-7+3\alpha(2-\alpha)\right]}.
\label{kappa'NS}
\eeqa

The presence of the term $d-4+3d\al$ in the denominators of Eqs.\ \eqref{kappaNS} and \eqref{muNS} implies that the heat flux transport coefficients $\kappa_\NS$ and $\mu_\NS$ \textit{diverge} when $\alpha$ tends (from above) to $\alpha_0=\frac{1}{3}$ and $\alpha_0=\frac{1}{9}$ for $d=2$ and $d=3$, respectively \cite{BGM10,S03}. However, the coefficient $\kappa'_\NS$ is finite for any $\alpha$ and any $d>1$.

The one-dimensional case deserves some care. As is known, the thermal conductivity in the elastic limit, $\kappa_0$, diverges at $d=1$ \cite{NR02}. Surprisingly enough, the thermal conductivity is well defined at $d=1$ for inelastic collisions ($\alpha<1$). Taking the limit $d\to 1$ in Eq.\ \eqref{63}  one gets $\kappa_\NS=(18p/m\nu)/(1-\alpha^2)$. On the other hand, the coefficient $\mu_\NS$ vanishes at $\alpha=1$ but diverges for $\alpha<1$ if $d=1$.

\begin{figure}[htbp]
\centerline{\includegraphics[width= .5\columnwidth]{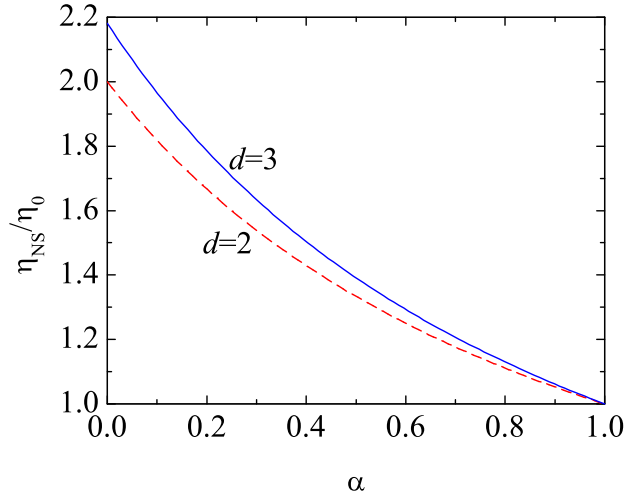}}
\caption{Plot of the reduced NS shear viscosity $\eta_\NS/\eta_0$ for $d=2$ (dashed
line) and $d=3$ (solid line).\label{fig2}}
\end{figure}

\begin{figure}[htbp]
\centerline{\includegraphics[width= .5\columnwidth]{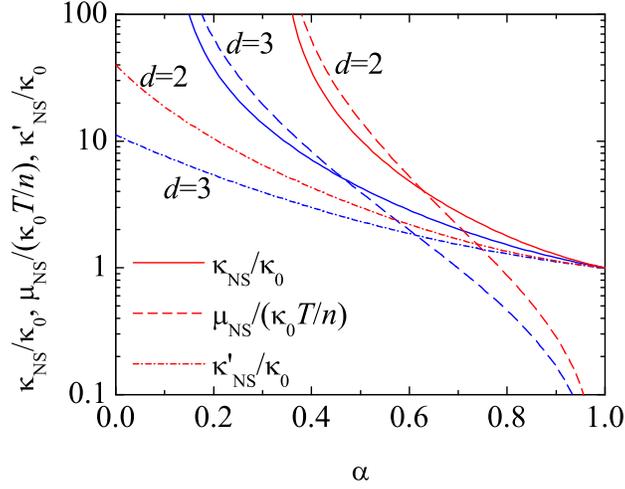}}
\caption{Plot of the reduced NS heat flux coefficients $\kappa_\NS/\kappa_0$ (solid lines), $\mu_\NS/(\kappa_0T/n)$ (dashed lines), and $\kappa'_\NS/\kappa_0$ (dash-dotted lines) for $d=2$  and $d=3$. The quantities $\kappa_\NS$ and $\mu_\NS$ diverge at $\alpha_0=\frac{1}{3}$ ($d=2$) and $\alpha_0=\frac{1}{9}$ ($d=3$).\label{fig3}}
\end{figure}

\begin{figure}[htbp]
\centerline{\includegraphics[width= .5\columnwidth]{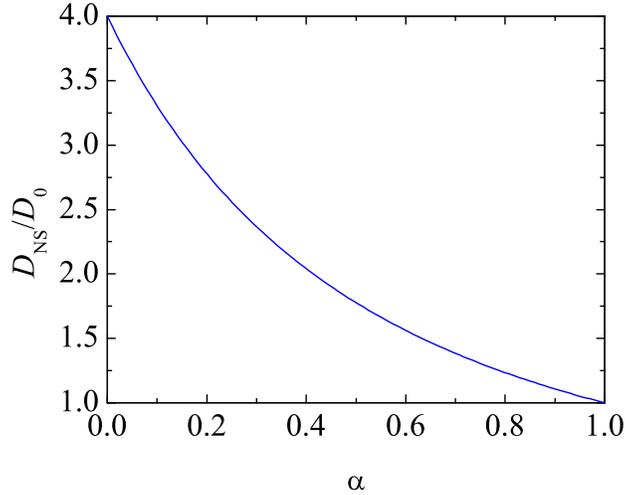}}
\caption{Plot of the reduced NS self-diffusion coefficient $D_\NS/D_0$ for any dimensionality $d$.\label{fig4}}
\end{figure}

Let us try to understand the origin of the divergence of $\kappa_\NS$ and $\mu_\NS$ when $\alpha\to\alpha_0$ for $d=2$ and $3$. The integral equation \eqref{c12} suggests that its solution has the form
\beq
f^\one(\mathbf{V})=\mathcal{A}(V)\mathbf{V}\cdot \nabla T+\mathcal{B}(V)\mathbf{V}\cdot \nabla n+\mathcal{C}(V)\left(V_i V_j-\frac{1}{d}V^2\delta_{ij}\right)\nabla_i u_j,
\label{f1}
\eeq
where $\mathcal{A}(V)$, $\mathcal{B}(V)$, and $\mathcal{C}(V)$ are unknown functions that only depend on the magnitude of velocity. The solvability conditions of Eq.\ \eqref{c12} imply that
\beq
\int\dd\mathbf{V}\, V^2\mathcal{A}(V)=\int\dd\mathbf{V}\, V^2\mathcal{B}(V)=0.
\eeq
The transport coefficients are directly related to velocity integrals of $\mathcal{A}(V)$, $\mathcal{B}(V)$, and $\mathcal{C}(V)$. Specifically,
\beq
\kappa_\NS=-\frac{m}{2d}\int\dd\mathbf{V}\, V^4\mathcal{A}(V),
\label{X1}
\eeq
\beq
\mu_\NS=-\frac{m}{2d}\int\dd\mathbf{V}\, V^4\mathcal{B}(V),
\label{X2}
\eeq
\beq
\eta_\NS=-\frac{m}{d(d+2)}\int\dd\mathbf{V}\, V^4\mathcal{C}(V).
\label{X3}
\eeq
The corresponding expression for the modified thermal conductivity coefficient $\kappa'_\NS$ is analogous to Eq.\ \eqref{X1}, except for the replacement $\mathcal{A}(\mathbf{V})\to \mathcal{A}'(\mathbf{V})\equiv \mathcal{A}(\mathbf{V})-\mathcal{B}(\mathbf{V})(n/2T)$.
Equations \eqref{f1}--\eqref{X3} are formally valid for both IMM and IHS. In the former case, however, the algebraic high-velocity tail $f^\zero\sim V^{-d-s(\alpha)}$ [cf.\ Eq.\ \eqref{18}] implies, according to Eqs.\ \eqref{48}--\eqref{50}, that $\mathbf{A}\sim \mathbf{B}\sim V^{-d-s(\alpha)+1}$ and $\mathsf{C}\sim V^{-d-s(\alpha)}$. One could therefore expect that the unknown functions defining $f^\one$ also present algebraic tails of the form $\mathcal{A}\sim \mathcal{B}\sim V^{-d-a(\alpha)}$, $\mathcal{A}'\sim  V^{-d-a'(\alpha)}$, and $\mathcal{C}\sim V^{-d-c(\alpha)}$. The convergence of $\eta_\NS$ and of $\kappa'_\NS$ implies that $c(\alpha)>4$ and $a'(\alpha)>4$ for all $\alpha$ and $d$. However, the divergence of $\kappa_\NS$ and $\mu_\NS$ at $\alpha=\alpha_0$ leads to $a(\alpha)\leq 4$  for $\alpha\leq \alpha_0$ if $d=2$ or $d=3$. This means that, although $f^\one(\mathbf{V})$ is well defined for any $\alpha$, its third-order velocity moments (such as the heat flux) might diverge due to the high-velocity tail of the distribution. This singular behavior is closely tied to the peculiarities of the IMM since the high-velocity tail of $f^\zero$ in the case of IHS is exponential \cite{BRMC96,EP97,NE98} rather than algebraic.

To close the evaluation of the NS transport coefficients, we now consider the self-diffusion coefficient defined by Eq.\ \eqref{Fick}. It is given by \cite{GA05}
\beqa
D_\NS&=&\frac{p}{\nuzo-\frac{1}{2}\zeta}\nn
&=&D_0\frac{4}{(1+\alpha)^2},
\eeqa
where $D_0=dp/\nu$ is the self-diffusion coefficient in the elastic limit and we have made use of Eqs.\ \eqref{10} and \eqref{nuzo} in the last step. In contrast to the other transport coefficients, the reduced self-diffusion coefficient $D_\NS/D_0$ is independent of the dimensionality of the system.

Figures \ref{fig2}--\ref{fig4} depict the $\alpha$-dependence of the reduced NS transport coefficients $\eta_\NS/\eta_0$, $\kappa_\NS/\kappa_0$, $\mu_\NS/(\kappa_0T/n)$, $\kappa'_\NS/\kappa_0$, and $D_\NS/D_0$ for $d=2$ and $d=3$. All of them increase with increasing dissipation. As for the influence of $d$, it depends on the transport coefficient under consideration. While, at a given value of $\alpha$, the shear viscosity increases with dimensionality, the opposite happens for the heat flux coefficients. This is especially apparent in the cases of $\kappa_\NS$ and $\mu_\NS$  since their divergence occurs at a smaller value $\alpha=\alpha_0$ for $d=2$ than for $d=3$. Finally, as said above, the reduced self-diffusion coefficient is independent of the dimensionality.
It is noteworthy that the coefficient $\mu_\NS/(\kappa_0T/n)$, which vanishes in the elastic case, becomes larger than $\kappa_\NS/\kappa_0$ for sufficiently high inelasticity.

\vspace*{0.5cm}
\setcounter{equation}{0}

\section{Uniform shear flow\label{sec5}}

The hydrodynamic description in the preceding section applies to arbitrary degree of dissipation provided that the hydrodynamic gradients are weak enough to allow for a NS theory.
The Chapman-–Enskog method assumes that
the relative changes of the hydrodynamic fields over distances on the order
of the mean free path are small. In the case of ordinary fluids this can be
controlled by the initial or boundary conditions.
For granular gases the situation
is more complicated, especially in  steady states, since there might be a relationship
between dissipation and gradients such that both cannot be chosen independently.
In spite of the above cautions, the NS approximation is appropriate in some important problems, such as
spatial perturbations of the HCS for an isolated system and the linear stability analysis of this state.
In this section and the two next ones we obtain in an exact way some  non-Newtonian hydrodynamic properties of a sheared granular gas modeled by the IMM.

The simple or uniform shear flow (USF) state is one of the most widely studied states, both for ordinary \cite{GS03} and granular gases \cite{C90,Go03}.
It is characterized by a constant density $n$, a uniform
granular temperature $T$, and a linear velocity profile $u_x=a y$,
where $a$ is the constant shear rate. At
a microscopic level, the USF is characterized by a velocity
distribution function that becomes \textit{uniform} in the local
Lagrangian frame, i.e.,
\beq
f(\mathbf{r},\mathbf{v};t)=f(\mathbf{V},t).
\eeq
In this frame, the Boltzmann equation \eqref{1a} reduces to
\beq
\partial_t f(\mathbf{V})-aV_y\frac{\partial}{\partial
V_x}f(\mathbf{V})=J[\mathbf{V}|f,f].
\label{W1}
\eeq
Equation \eqref{W1} is invariant under the transformations
\beq
(V_x,V_y)\to (-V_x,-V_y),
\label{VxVy}
\eeq
\beq
V_j\to -V_j,\quad V_j\to V_k,\quad j,k\neq x,y.
\label{Vj}
\eeq
This implies that if the initial state $f(\mathbf{V},0)$ is
consistent with the symmetry properties \eqref{VxVy} and \eqref{Vj}
so is the solution to Eq.\ \eqref{W1} at any time $t>0$. Even if one
starts from an initial condition inconsistent with \eqref{VxVy} and
\eqref{Vj}, it is expected that the solution asymptotically tends
for long times to a function compatible with \eqref{VxVy} and
\eqref{Vj}.

The properties of uniform temperature and constant density and shear
rate are enforced in computer simulations by applying the
Lees--Edwards boundary conditions \cite{GS03,LE72}, regardless of
the particular interaction model considered. In the case of boundary
conditions representing realistic plates in relative motion, the
corresponding nonequilibrium state is the so-called Couette flow,
where density, temperature, and shear rate are no longer uniform
\cite{TTMGSD01}.

According to the conditions defining the USF, the balance equations \eqref{b7} and \eqref{b8} are satisfied identically, while Eq.\ \eqref{b9} becomes
\beq
\partial_t T=- \frac{2}{dn}P_{xy} a-\zeta T.
\label{5.1}
\eeq
This balance equation shows that the temperature changes in
time due to two competing effects: the viscous heating term $-P_{xy}a$ and the
inelastic collisional cooling term $\zeta T$. Depending on the initial condition,
one of the effects prevails over the other one so that the
temperature either increases or decreases in time. Eventually, a steady
state is reached for sufficiently long times when both effects cancel each other. In this steady state
\beq
-\frac{P_{xy}}{p}=\frac{d}{2} \frac{\zeta}{a}.
\label{5.2}
\eeq
This relation illustrates the connection between inelasticity (as measured by the cooling rate $\zeta$), irreversible fluxes (as measured by the shear stress $P_{xy}/p$), and hydrodynamic gradients (as measured by the shear rate $a$).

The rheological properties are related to the non-zero elements of the  pressure tensor consistent with Eqs.\ \eqref{VxVy} and \eqref{Vj}, namely $P_{xy}$, $P_{xx}$, $P_{yy}$, and $p$. The remaining $d-2$ diagonal elements are equal, by symmetry, so that $P_{zz}=\cdots=P_{dd}=(dp-P_{xx}-P_{yy})/(d-2)$. In order to obtain these four independent elements, we complement Eq.\ \eqref{5.1} with the equations obtained by multiplying both sides of Eq.\ \eqref{W1} by $\{V_xV_y,V_x^2,V_y^2\}$ and integrating over velocity. The result is
\beq
\partial_t P_{xy}+aP_{yy}=-\nuzt P_{xy},
\label{5.3}
\eeq
\beq
\partial_t P_{xx}+2aP_{xy}=-\nuzt \left(P_{xx}-p\right)-\zeta p,
\label{5.3b}
\eeq
\beq
\partial_t P_{yy}=-\nuzt \left(P_{yy}-p\right)-\zeta p,
\label{5.4}
\eeq
where we have taken into account Eq.\ \eqref{Z1}.

\subsection{Steady-state solution\label{sec5a}}

The \textit{steady-state} solution of Eq.\ \eqref{5.4} is simply
\beqa
P_{yy}^*&=&1-\frac{\zeta}{\nuzt}\nn
&=&\frac{d}{2}\frac{1+\alpha}{d+1-\alpha}.
\label{5.5}
\eeqa
Here we have introduced the reduced pressure tensor $P_{ij}^*=P_{ij}/p$.
Substitution into Eq.\ \eqref{5.3} yields, again in the steady state,
\beqa
P_{xy}^*&=&-\frac{a}{\nuzt}\left(1-\frac{\zeta}{\nuzt}\right)\nn
&=&-\frac{a}{\nu}\frac{d^2(d+2)}{2(d+1-\alpha)^2}.
\label{5.6}
\eeqa
Next,  Eq.\ \eqref{5.3b} gives in the steady state
\beqa
P_{xx}^*&=&1-\frac{\zeta}{\nuzt}+2\frac{a^2}{\nuzt^2}\left(1-\frac{\zeta}{\nuzt}\right)\nn
&=&\frac{d}{2}\frac{1+\alpha}{d+1-\alpha}+\frac{a^2}{\nu^2}\frac{d^3(d+2)^2}{(d+1-\alpha)^3(1+\alpha)}.
\label{5.9}
\eeqa
Equations \eqref{5.6} and \eqref{5.9} are not closed  since the ratio $\nu/a$, which yields the steady-state temperature for a given shear rate $a$,  must be determined. This is done by elimination of $P_{xy}^*$ between Eqs.\ \eqref{5.2} and \eqref{5.6} with the result
\beqa
\frac{\nu^2}{a^2}&=&\frac{2}{d}\frac{\nu^2}{\nuzt\zeta}\left(1-\frac{\zeta}{\nuzt}\right)\nn
&=&\frac{2d^2(d+2)}{(1-\alpha^2)(d+1-\alpha)^2}.
\label{5.7}
\eeqa
Using this result in Eqs.\ \eqref{5.6} and \eqref{5.9}, we obtain the $\alpha$-dependence of $P_{xy}^*$ and $P_{xx}^*$:
\beq
P_{xy}^*=-\frac{d\sqrt{(d+2)(1-\alpha^2)}}{2\sqrt{2}(d+1-\alpha)}.
\label{5.8}
\eeq
\beq
P_{xx}^*=\frac{d}{2}\frac{d+3-(d+1)\alpha}{d+1-\alpha}.
\label{5.10}
\eeq

Equations \eqref{5.5}, \eqref{5.8}, and \eqref{5.10} provide the explicit expressions of the relevant elements of the reduced pressure tensor as functions of the coefficient of restitution $\alpha$ and the dimensionality $d$. Since $\nu^2\propto T$, Eq.\ \eqref{5.7} shows that the steady-state temperature is proportional to the square of the shear rate. Equation \eqref{5.7} can also be interpreted as expressing the \textit{reduced} shear rate $a/\nu$ as a function of $\alpha$. As a consequence, no matter how large or small the shear rate $a$ is, its strength relative to the stationary collision frequency $\nu$ is fixed by the value of $\alpha$, so one cannot choose the steady-state value of $a/\nu$ independently of $\alpha$. It is important to notice that, according to Eqs.\ \eqref{5.5} and \eqref{5.10}, $P_{xx}^*+(d-1)P_{yy}^*=d$. This implies that $P_{zz}^*=P_{yy}^*$, even though the directions $y$ and $z$ are physically different in the geometry of the USF.

In order to characterize the rheological properties in the USF, it is convenient to introduce a generalized shear viscosity $\eta$ and a (first) viscometric function $\Psi$ by
\beq
P_{xy}=-\eta\frac{\partial u_x}{\partial y},
\label{5.11}
\eeq
\beq
P_{xx}-P_{yy}=\Psi\left(\frac{\partial u_x}{\partial y}\right)^2.
\label{5.12}
\eeq
The second viscometric function vanishes as a consequence of the property $P_{zz}=P_{yy}$. {}From Eq.\ \eqref{5.6} one obtains
\beq
\eta=\eta_0\left(\frac{d}{d+1-\alpha}\right)^2,
\label{5.13}
\eeq
where we recall that $\eta_0=(d+2)(p/2\nu)=p/\nuhs$ is the NS shear viscosity at $\alpha=1$. Analogously, Eqs.\ \eqref{5.5} and \eqref{5.9} yield
\beq
\Psi=\Psi_0\frac{2}{1+\alpha}\left(\frac{d}{d+1-\alpha}\right)^3,
\label{5.14}
\eeq
where $\Psi_0=2{\eta_0^2}/{p}$ is the corresponding Burnett coefficient in the elastic limit \cite{CC70}.

\begin{figure}[htbp]
\centerline{\includegraphics[width= .5\columnwidth]{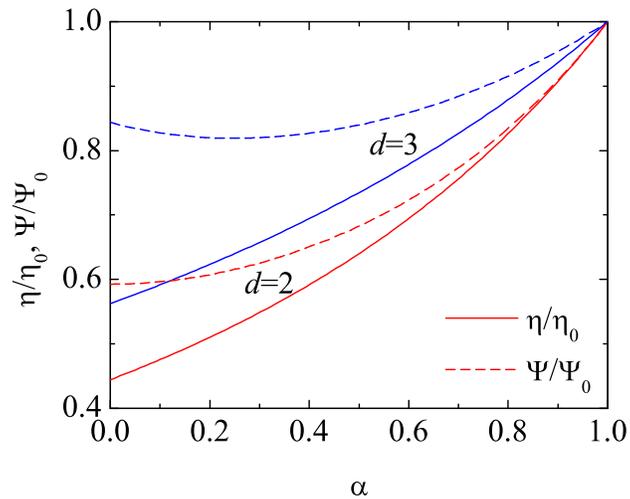}}
\caption{Plot of the reduced rheological functions $\eta/\eta_0$ (solid lines) and $\Psi/\Psi_0$ (dashed lines) in the steady-state USF for $d=2$ and $d=3$.\label{fig5}}
\end{figure}

\begin{figure}[htbp]
\centerline{\includegraphics[width= .5\columnwidth]{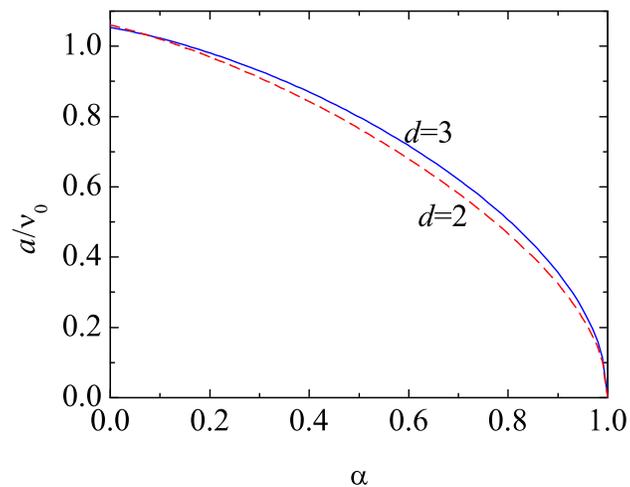}}
\caption{Plot of the reduced shear rate $a/\nuhs$  in the steady-state USF for $d=2$ (dashed line) and $d=3$ (solid line).\label{fig6}}
\end{figure}

\begin{figure}[htbp]
\centerline{\includegraphics[width= .5\columnwidth]{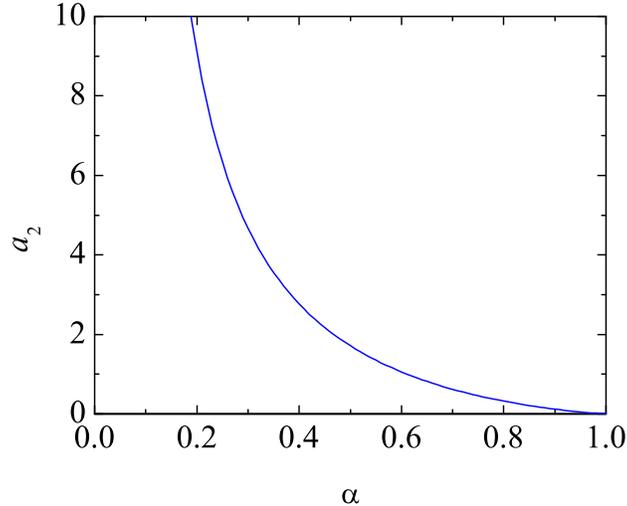}}
\caption{Plot of the fourth cumulant $a_2$  in the steady-state USF for  $d=3$. This quantity diverges at $\alpha_c\simeq 0.046$.\label{fig7}}
\end{figure}

Figure \ref{fig5} shows the $\alpha$-dependence of the rheological quantities $\eta/\eta_0$ and $\Psi/\Psi_0$ for $d=2$ and $d=3$. It is apparent that $\eta/\eta_0$ is a monotonically decreasing function of inelasticity, this effect being more pronounced in the two-dimensional case than in the three-dimensional one. This decrease contrasts dramatically with the behavior of the NS shear viscosity, as seen in Fig.\ \ref{fig2}. This confirms that the transport properties in the steady-state USF are inherently different from those of the NS description \cite{SGD04}. Another non-Newtonian feature is the existence of normal stress differences in the shear flow plane. What is interesting is that the viscometric coefficient $\Psi$ measuring this effect strongly deviates (in general) from its elastic Burnett-order value $\Psi_0$. We observe from Fig.\ \ref{fig5} that this effect is again more significant for $d=2$ than for $d=3$. Moreover, in the former case  $\Psi/\Psi_0$ monotonically decreases with decreasing $\alpha$, while it reaches a minimum at  $\alpha=0.25$ in the three-dimensional case. To complement this discussion, it is worth plotting the steady-state reduced shear rate $a^*\equiv a/\nuhs$ versus $\alpha$. This is done in Fig.\ \ref{fig6} for $d=2$ and $d=3$. Since $a=\partial u_x/\partial y$ is the only gradient present in the USF, the ratio $a/\nuhs$ measures the \textit{relative} strength of the hydrodynamic gradients and thus the departure from the homogeneous state. Therefore, it plays the role of the Knudsen number. Figure \ref{fig6} shows that $a^*$ increases with inelasticity, having an infinite slope at $\alpha=1$. The influence of dimensionality on this quantity is much weaker that in the cases of $\eta/\eta_0$ and $\Psi/\Psi_0$.

Although all the previous results in this section are exactly derived from the Boltzmann equation \eqref{W1} for IMM, the solution to this equation is not known. However, we can get some indirect information about the distribution function $f(\mathbf{V})$ through its moments.
In principle, the hierarchy of moment equations stemming from Eq.\ \eqref{W1} can be recursively solved since the equations for moments of order $k$ involve only  moments of the same and lower order.
Equations \eqref{5.5} and \eqref{5.9}--\eqref{5.10} give the second-order moments. The next non trivial moments are of fourth-order. They were obtained (for $d=3$) in Ref.\ \cite{SG07} as the solution of a set of eight linear, inhomogeneous equations. The results show that the fourth-order moments are finite for $\alpha>\alpha_c$, where $\alpha_c\simeq 0.046$ is a critical value below which the fourth-order moments diverge. This implies that the distribution function exhibits a high-energy tail of the form $f(\mathbf{V})\sim V^{-d-s'(\alpha)}$, so that $s'(\alpha)<4$ if $\alpha<\alpha_c$ for $d=3$. This tail in the USF is reminiscent of that of the HCS [see Eq.\ \eqref{18}]. On the other hand, the fourth-order moments are finite in the HCS for $d\geq 2$ and any value of $\alpha$ [see Eq.\ \eqref{a2B}]. This suggests that $s'(\alpha)<s(\alpha)$, i.e., the shearing enhances the overpopulation of the high-velocity tail. As an illustration, Fig.\ \ref{fig7} displays the fourth cumulant $a_2$ of the USF, defined by Eq.\ \eqref{a2A}, as a function of $\alpha$ for $d=3$. Comparison with Fig.\ \ref{fig1} shows that this quantity is much larger in the USF than in the HCS.

\subsection{Unsteady hydrodynamic solution\label{sec5b}}
The interest of the USF is not restricted to the steady state. In general, starting from an arbitrary initial temperature $T(0)$, the temperature $T(t)$ changes in time according to Eq.\ \eqref{5.1} either by increasing (if the viscous heating term dominates over the collisional cooling term) or decreasing (in the opposite case). After a short kinetic stage (of the order of a few mean free times) and before reaching the steady state, the system follows an  unsteady \textit{hydrodynamic} regime where the reduced pressure tensor $P_{ij}^*(t)$ depends on time through a dependence on the reduced shear rate $a^*(t)=a/\nu_0(t)$, in such a way that the functions $P_{ij}^*(a^*)$ are independent of the initial condition \cite{AS07,SGD04}.

Taking into account that $\nu_0\propto T^{1/2}$, one has
\beqa
\partial_t a^*&=&-\frac{a^*}{2T}\partial_t T\nn
&=&\frac{a^*}{2}\left(\zeta+\frac{2a}{d} P_{xy}^*\right),
\label{5.15}
\eeqa
where in the last step use has been made of Eq.\ \eqref{5.1}. As a consequence,
\beqa
\partial_t P_{ij}&=&P_{ij}^*\partial_t p+p\left(\partial_{a^*}P_{ij}^*\right)\partial_t a^*\nn
&=&-p\left(\zeta+\frac{2a}{d} P_{xy}^*\right)\left(1-\frac{a^*}{2}\partial_{a^*}\right)P_{ij}^*.
\label{5.16}
\eeqa
Insertion of this property into Eqs.\ \eqref{5.3}--\eqref{5.4} yields
\beq
\partial_{a^*}P_{xy}^*=\frac{2}{a^*}\left(P_{xy}^*-\frac{\nuzt^* P_{xy}^*+a^* P_{yy}^*}{\zeta^*+\frac{2a^*}{d} P_{xy}^*}\right),
\label{5.17}
\eeq
\beq
\partial_{a^*}P_{yy}^*=\frac{2}{a^*}\left[P_{yy}^*-\frac{\nuzt^* \left(P_{yy}^*-1\right)+\zeta^*}{\zeta^*+\frac{2a^*}{d} P_{xy}^*}\right],
\label{5.18}
\eeq
\beq
\partial_{a^*}P_{xx}^*=\frac{2}{a^*}\left[P_{xx}^*-\frac{\nuzt^* \left(P_{xx}^*-1\right)+\zeta^*+2a^*P_{xy}^*}{\zeta^*+\frac{2a^*}{d} P_{xy}^*}\right],
\label{5.19}
\eeq
where we have called $\nuzt^*\equiv \nuzt/\nuhs$ and $\zeta^*\equiv \zeta/\nuhs$.

\begin{figure}[h]
\centerline{\includegraphics[width= .5\columnwidth]{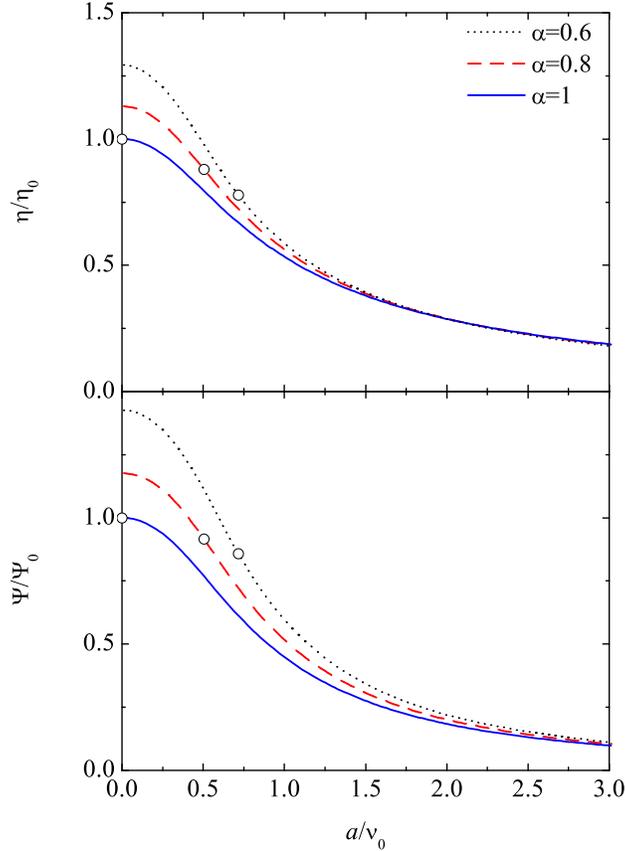}}
\caption{Plot of the reduced generalized shear viscosity $\eta/\eta_0$ (top panel) and of the reduced generalized viscometric function $\Psi/\Psi_0$ (bottom panel) versus the reduced shear rate $a/\nu_0$ in the unsteady USF for $d=3$ and three coefficients of restitution: $\alpha=0.6$ (dotted curves), $\alpha=0.8$ (dashed lines), and $\alpha=1$ (solid lines). The circles denote the steady-state values.\label{fig7b}}
\end{figure}

Equations \eqref{5.17} and \eqref{5.18} constitute a set of two coupled  nonlinear first-order differential equations for the elements $P_{xy}^*$ and $P_{yy}^*$. Their numerical solution, with appropriate boundary conditions \cite{SGD04}, provides the hydrodynamic functions $P_{xy}^*(a^*)$ and $P_{yy}^*(a^*)$. Moreover, it is straightforward to check that Eqs.\ \eqref{5.18} and \eqref{5.19} are consistent with the relationship $P_{xx}^*+(d-1)P_{yy}^*=d$. This means that the knowledge of $P_{yy}^*(a^*)$ suffices to determine $P_{xx}^*(a^*)$ and that $P_{zz}^*(a^*)=P_{yy}^*(a^*)$. This generalizes the analogous relations (in particular, a vanishing second viscometric function) obtained above in the steady state. {}Once  $P_{xy}^*(a^*)$ and $P_{yy}^*(a^*)$ are known, the shear-rate dependence of the generalized shear viscosity $\eta$ and (first) viscometric function $\Psi$, defined by Eqs.\ \eqref{5.11} and \eqref{5.12}, can be obtained. These two functions are plotted in Fig.\ \ref{fig7b} for $\alpha=0.6$, $\alpha=0.8$, and $\alpha=1$ in the three-dimensional case. The top panel clearly shows that  the shear viscosity exhibits shear thinning, i.e., it decays with increasing reduced shear rate. As $a^*$ increases the influence of inelasticity on the shear viscosity becomes less important.  The top panel of Fig.\ \ref{fig7b} also shows that the NS value (i.e., the value at $a^*=0$) of the shear viscosity increases with increasing inelasticity, in agreement with Fig.\ \ref{fig2}. On the other hand, the steady-state values (which correspond to different values of $a^*$) decrease as inelasticity increases, in agreement with Fig.\ \ref{fig5}.
Analogous features are presented by the viscometric function plotted in the bottom panel.

Although the determination of $P_{xy}^*(a^*)$ and $P_{yy}^*(a^*)$ involves numerical work, one can obtain analytically those functions in the vicinity of the steady state by means of the derivatives
$\partial_{a^*}P_{xy}^*$ and $\partial_{a^*}P_{yy}^*$ evaluated at the steady state. This requires some care because the fractions on the right-hand side of Eqs.\ \eqref{5.17} and \eqref{5.18} become indeterminate  in the steady state since the numerators and the denominator vanish identically. This difficulty can be solved by means of L'H\^opital rule \cite{G06}. Therefore, in the steady-state limit Eqs.\ \eqref{5.17} and \eqref{5.18} become
\beq
\partial_{a^*}P_{xy}^*=\frac{2}{a^*}\left(P_{xy}^*-\frac{d}{2}\frac{\nuzt^* \partial_{a^*}P_{xy}^*+ P_{yy}^*+a^*\partial_{a^*}P_{yy}^*}{P_{xy}^*+a^*\partial_{a^*}P_{xy}^*}\right),
\label{5.20}
\eeq
\beq
\partial_{a^*}P_{yy}^*=\frac{2}{a^*}\left(P_{yy}^*-\frac{d}{2}\frac{\nuzt^* \partial_{a^*}P_{yy}^*}{P_{xy}^*+a^*\partial_{a^*}P_{xy}^*}\right).
\label{5.21}
\eeq
Elimination of $\partial_{a^*}P_{yy}^*$ gives a cubic equation for $\partial_{a^*}P_{xy}^*$,
\beqa
&&{a^*}^3 \left(\partial_{a^*}P_{xy}^*\right)^3+2d\nuzt^* a^*\left(\partial_{a^*}P_{xy}^*\right)^2+\left[d^2\frac{{\nuzt^*}^2}{a^*}+3a^*\left(d P_{yy}^*-{P_{xy}^*}^2\right)\right]
\partial_{a^*}P_{xy}^*\nn
&&+2dP_{xy}^*P_{yy}^*+\left(d P_{yy}^*-2{P_{xy}^*}^2\right)\left(P_{xy}^*+d\frac{\nuzt^*}{a^*}\right)=0,
\label{5.22}
\eeqa
with coefficients that are known functions of $\al$. The real root of Eq.\ \eqref{5.22} gives the physical solution. {}From it we simply get
\beq
\partial_{a^*}P_{yy}^*=2P_{yy}^*\frac{P_{xy}^*+a^*\partial_{a^*}P_{xy}^*}{a^*P_{xy}^*+d\nuzt^*+{a^*}^2\partial_{a^*}P_{xy}^*}.
\label{5.23}
\eeq

\vspace*{0.5cm}
\setcounter{equation}{0}

\section{Couette flow with uniform heat flux. LTu flow\label{sec6}}

The planar Couette flow  corresponds to a
granular gas enclosed between two parallel, infinite plates (normal to the $y$ axis) in relative motion along the $x$ direction, and kept at different temperatures. The resulting flow velocity is along the $x$ axis and, from symmetry, it is expected that the hydrodynamic fields only vary in the $y$ direction.

Despite the apparent similarity between the steady planar Couette flow and the USF, the former  is much more complex than the latter. In contrast to the USF, the temperature is not uniform and thus a heat flux vector $\mathbf{q}$ coexists with the pressure tensor $P_{ij}$ \cite{TTMGSD01}. In general, inelastic cooling and viscous heating are unbalanced, their difference dictating the sign of the divergence of the heat flux \cite{TTMGSD01,VU09}. More explicitly, the energy balance equation \eqref{b9} in the steady state reads
\begin{equation}
-\frac{\partial q_y}{\partial y}=\frac{d}{2}\zeta nT+P_{xy}a,
\label{Tbal}
\end{equation}
where  we have again called $a\equiv {\partial u_x}/{\partial y}$. However, in contrast to the USF, the shear rate $a$ is not uniform, i.e., the velocity profile is not linear.
Conservation of momentum implies [see Eq.\ \eqref{b8}] $P_{xy}=\text{const}$ and $P_{yy}=\text{const}$.

The key difference between the balance equations \eqref{5.2} and \eqref{Tbal} is the presence of the divergence of the heat flux in the latter. Therefore, Eq.\ \eqref{Tbal} reduces to Eq.\ \eqref{5.2} if $\partial_y q_y=0$, even if $q_y\neq 0$ and $\partial_y T\neq 0$. This yields a whole new set of steady states where an exact balance between the viscous heating term and the collisional cooling term occurs at all points of the system. This class of Couette flows has been observed in computer simulations of IHS  and studied theoretically by means of Grad's approximate method and a simple kinetic model \cite{VGS10,VSG10}. Interestingly, an \textit{exact} solution of the Boltzmann equation for IMM supports this class of Couette flows with uniform heat flux \cite{SGV10}.

In the geometry of the planar Couette flow, the Boltzmann equation \eqref{1a} becomes
\beq
v_y\partial_s f=\frac{1}{\omegazt}J[f,f],
\label{V3}
\eeq
where we have particularized to the steady state and have introduced the scaled variable $s$ as
\beq
\dd s=\omegazt \dd y,
\label{V2}
\eeq
where
\beq
\omegazt\equiv\nuzt-\zeta=\frac{(1+\alpha)^2}{2(d+2)}\nu.
\eeq
An exact \textit{normal} solution of Eq.\ \eqref{V3} exists characterized by the following hydrodynamic profiles \cite{SGV10}:
\beq
p=\text{const},\quad \frac{\partial u_x}{\partial s}=\at=\text{const},\quad \frac{\partial T}{\partial s}=\text{const}.
\label{V4}
\eeq
Note that $\at=a/\omegazt\propto a/\nu$. Thus, it is of the order of the Knudsen number associated with the shear rate. It is important to bear in mind that, since $\widetilde{a}=\text{const}$, the ratio $a/\nu$ is spatially uniform even though neither the shear rate $a$ nor the collision frequency $\nu$ are.  Apart from $\at$, there is another Knudsen number, this time associated with the thermal gradient. It can be conveniently defined as
\beq
\et=\sqrt{2T/m}\frac{\partial \ln T}{\partial s}.
\label{V5}
\eeq
This quantity is not constant since $\partial_s T=\text{const}$ implies $\et\propto T^{-1/2}$.
As will be seen, the consistency of the profiles \eqref{V4} is possible only if $\at$ takes a certain particular value for each coefficient of restitution $\al$. In contrast, the reduced thermal gradient $\et$ will remain free and so independent of $\al$ \cite{SGV10}.

{}From Eqs.\ \eqref{V4} and \eqref{V5} we get
\beq
\frac{\partial T}{\partial u_x}=\frac{\et\sqrt{mT/2}}{\at}=\text{const}.
\label{LTu}
\eeq
This means that when the spatial variable ($y$ or $s$) is eliminated to express $T$ as a function of $u_x$ one gets a linear relationship. For this reason, the class of states defined by Eq.\ \eqref{V4} is referred to as the LTu class \cite{SGV10,SGV10,VSG10}.
In terms of the  variable $s$, the temperature profile is
\beq
T(s)=T_0\left(1+\frac{\et_0}{v_0}s\right),\quad v_0\equiv\sqrt{2T_0/m},
\label{V19}
\eeq
where $T_0$ and $\et_0$ are the temperature and Knudsen number at a reference point $s=0$.

In order to get the pressure tensor and the heat flux in the LTu flow, it is convenient to introduce the dimensionless velocity distribution function
\beq
\phi(\mathbf{c};\et)=\frac{T(s)}{p}\left[\frac{2T(s)}{m}\right]^{d/2}f(s,\mathbf{v}),\quad \mathbf{c}=\frac{\mathbf{v}-\mathbf{u}(s)}{\sqrt{2T(s)/m}}.
\label{V6}
\eeq
As a normal solution, all the dependence of $f$ on $s$ must occur through the hydrodynamic fields $T$ and $u_x$. Consequently,
\beq
\frac{\partial f}{\partial s}=\frac{\partial T}{\partial s}\frac{\partial f}{\partial T}+ \frac{\partial u_x}{\partial s}\frac{\partial f}{\partial u_x}.
\label{V7}
\eeq
Taking into account Eq.\ \eqref{V6}, and after some algebra, one gets \cite{SGV10}
\beqa
-c_y\left[\frac{\et}{2}\left(2+
\frac{\partial}{\partial\mathbf{c}}\cdot\mathbf{c}+\et\frac{\partial}{\partial
\et}\right)+\at\frac{\partial}{\partial c_x}\right]\phi&=&\frac{2(d+2)}{(1+\alpha)^2\Omega_d}\int
\dd{\bf c}_{1}\int \dd\widehat{\boldsymbol{\sigma}} \left[
\alpha^{-1}\phi({\bf c}')\phi({\bf c}_{1}')\right.\nn
&&\left.-\phi({\bf c})\phi({\bf c}_{1})\right] \nn
&\equiv &\mathcal{J}[\mathbf{c}|\phi,\phi].
\label{V11}
\eeqa

We consider now the following  moments of order $k=2r+\ell$,
\beq
\mathcal{M}_{2r|\ell,{h}}(\et)=\int \dd\mathbf{c}\,c^{2r}c_y^{\ell{-h}}{c_x^h}
\phi(\mathbf{c};\et),\quad 0\leq h\leq\ell.
\label{V12}
\eeq
By definition, $\mathcal{M}_{0|0,0}=1$, $\mathcal{M}_{0|1,0}=\mathcal{M}_{0|1,1}=0$, and $\mathcal{M}_{2|0,0}=\frac{d}{2}$. According to Eq.\ \eqref{V11}, the moment equations read
\beqa
\frac{\et}{2}\left(2r+\ell-1-\et\frac{\partial}{\partial
\et}\right)\mathcal{M}_{2r|\ell+1,{h}}{+\at\left(2r \mathcal{M}_{2r-2|\ell+2,h+1}+h\mathcal{M}_{2r|\ell,h-1}\right)}
&=&\mathcal{J}_{2r|\ell,{h}},
\label{V13}
\eeqa
where
\beq
\mathcal{J}_{2r|\ell,{h}}\equiv\int \dd\mathbf{c}\,c^{2r}c_y^{\ell{-h}}{c_x^h} \mathcal{J}[\mathbf{c}|\phi,\phi]
\eeq
are the corresponding collisional moments. In the case of IMM, $\mathcal{J}_{2r|\ell,{h}}$ is given as a bilinear combinations of the form  $\mathcal{M}_{2r'|\ell',{h'}}\mathcal{M}_{2r''|\ell'',{h''}}$ such that $2r'+\ell'+2r''+\ell''=2r+\ell$. Therefore, only moments of order  equal to or smaller than $2r+\ell$ contribute to $\mathcal{J}_{2r|\ell,{h}}$.

It can be verified that the hierarchy \eqref{V13} is consistent with solutions of the form \cite{SGV10}
\beq
\mathcal{M}_{2r|\ell, h}(\et)=\sum_{j=0}^{2r+\ell-2}\mu_j^{(2r|\ell, h)}\et^j,\quad \mu_j^{(2r|\ell, h)}=0\text{ if }j+\ell=\text{odd},
\label{V14}
\eeq
i.e., the moments $\mathcal{M}_{2r|\ell,{h}}(\et)$ of order $2r+\ell\geq 2$ are  \textit{polynomials} in the thermal Knudsen number $\et$ of degree $2r+\ell-2$ and parity $\ell$.

 It is important to remark that, when $\et=0$, the hierarchy \eqref{V13} reduces to that of the (stationary) USF problem for IMM, i.e., the hierarchy obtained from Eq.\ \eqref{W1}.
In other words, the USF moments provide the independent terms of the corresponding LTu moments. On the other hand, the hierarchy \eqref{V13}  reduces to that of the conventional Fourier flow problem for elastic Maxwell particles when $\al=1$ (which, as will be seen below, implies $\at=0$) \cite{GS03,S09}.
The general problem ($\et\neq 0$, $\at\neq 0$) is much more difficult since it combines both momentum and energy transport. The interesting point is that, although the  moment hierarchy  \eqref{V13} couples moments of order $k$ to moments of a higher order $k+1$, it can be exactly solved via a recursive scheme. This is possible because  the coefficient $\mu_j^{(2r|\ell+1,h)}$ with $j= 2r+\ell-1$ of the moment $\mathcal{M}_{2r|\ell+1,h}$  do not contribute to Eq.\ \eqref{V13}. In the following, we will focus on the moments of second order (pressure tensor) and third order (heat flux).

Since the moments of order $k$ are polynomials in $\et$ of degree $k-2$, it turns out that the elements of the pressure tensor are independent of the reduced thermal gradient $\et$. Therefore, they coincide with those obtained in the steady-state USF, being given by Eqs.\ \eqref{5.5}, \eqref{5.8}, and \eqref{5.10}. Moreover, the reduced shear rate $a/\nu$ is again given by Eq.\ \eqref{5.7}, so that
\beq
\at=\frac{d+1-\al}{d(1+\al)^2}\sqrt{2(d+2)(1-\al^2)}.
\eeq
This confirms that, as said above, the value of $\widetilde{a}$ in the LTu flow is enslaved to the coefficient of restitution $\al$. If one defines the rheological functions $\eta$ and $\Psi$ by Eqs.\ \eqref{5.11} and \eqref{5.12}, their expressions in the LTu flow are the same as those in the USF [see Eqs.\ \eqref{5.13} and \eqref{5.14}].

The  third-order moments (absent in the USF) are linear functions of $\et$ that cannot be evaluated autonomously. However, they can be obtained from Eq.\ \eqref{V13}  in terms of the independent terms of the fourth-order moments \cite{SGV10}. The explicit forms for the two third-order moments defining the $x$ and $y$ components of the heat flux are
\beqa
\mathcal{M}_{2|1,0}(\et)&=&-\frac{2d\et}{ X}\bigg\{
  4 d  (2d - 2 + 3d \zt) \zt \mu_0^{(0|4,0)}+
      8 d\zt \mu_0^{(0|4,2)} + (18 d -18+ 19d \zt)\mu_0^{(2|2,0)}\nn
      &&\ -
  6 \sqrt{2d\zt}  \left[ (2d - 2 + 3d \zt) \mu_0^{(0|4,1)} +
     \mu_0^{(2|2,1)}\right]\bigg\} ,
  \label{Z9}
       \eeqa
 \beqa
\mathcal{M}_{2|1,1}(\et)&=&\frac{2d\et}{3X}   \bigg\{\sqrt{2d\zt}\big[4 d (7d- 2  + 9 d \zt) \zt\mu_0^{(0|4,0)} +
 6  (6 d - 6 + 5 d\zt) \mu_0^{(0|4,2)}\nn&& +
 3 (9{d} + 6 + 17 d\zt) \mu_0^{(2|2,0)}\big]-12 d \zt  (7d - 2  + 9 d \zt) \mu_0^{(0|4,1)}\nn
 && -
 9  (6 d - 6 + 5d \zt) \mu_0^{(2|2,1)}\bigg\},
 \label{Z10}
       \eeqa
 where
 \beq
 X\equiv  36 (d - 1)^2 -  d\left(76 -56 d - 9  d\zt\right)\zt
       \label{Z11}
 \eeq
 and
\beq
\zt\equiv\frac{\nutz}{\omegazt}=\frac{d+2}{d}\frac{1-\al}{1+\al}.
\label{X6a}
\eeq
The coefficients $\mu_0^{(0|4,0)}$, $\mu_0^{(0|4,2)}$, $\mu_0^{(2|2,0)}$, $\mu_0^{(0|4,1)}$, and $\mu_0^{(2|2,1)}$ are not but the moments $\mathcal{M}_{0|4,0}$, $\mathcal{M}_{0|4,2}$, $\mathcal{M}_{2|2,0}$, $\mathcal{M}_{0|4,1}$, and $\mathcal{M}_{2|2,1}$, respectively, evaluated in the USF \cite{SG07}. This completes the determination of $\mathcal{M}_{2|1,0}$ and $\mathcal{M}_{2|1,1}$.

Once the non-zero components of the heat flux $q_y$ and $q_x$ are known, one can define an effective  thermal conductivity $\kappa$ and a  cross coefficient $\Phi$, respectively, by
\beq
q_y=-\kappa\frac{\partial T}{\partial y},\quad q_x=\Phi \frac{\partial T}{\partial y}\frac{\partial u_x}{\partial y}.
\label{Z14}
\eeq
In the three-dimensional case the expressions of $\kappa$ and $\Phi$ are
\beq
\kappa=\kappa_0\frac{36 }{  (4 - \al)^2 (829 - 162 \al - 91 \al^2)}\frac{A(\al)}{C(\al)},
\label{Z23}
\eeq
\beq
\Phi=\Phi_0\frac{864}{ 35(1+\al)  (4 - \al)^3 (829 - 162 \al - 91 \al^2)}\frac{B(\al)}{C(\al)},
\label{Z24}
\eeq
where the functions $A(\al)$, $B(\al)$, and $C(\al)$ are polynomials in $\al$ of degrees 26, 26, and 24, respectively, whose coefficients are given in Table 1 of Ref.\ \cite{SGV10}.
In Eq.\ \eqref{Z24}, $\Phi_0=\frac{7}{2}\kappa_0\eta_0/p$ is the corresponding Burnett coefficient in the elastic limit for $d=3$ \cite{CC70}.

\begin{figure}[htbp]
\centerline{\includegraphics[width= .5\columnwidth]{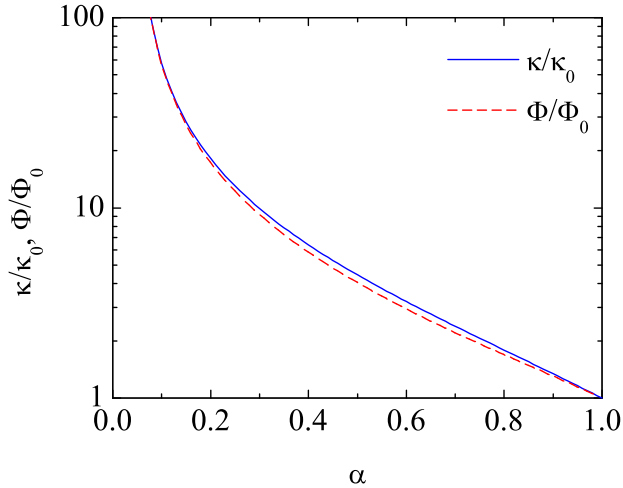}}
\caption{Plot of the reduced heat flux coefficients  $\kappa/\kappa_0$ (solid line) and $\Phi/\Phi_0$  (dashed line) in the LTu flow for $d=3$. These quantities diverge at $\alpha_c\simeq 0.046$.\label{fig8}}
\end{figure}

While the coefficient $\kappa$ is an extension of the conventional NS thermal conductivity, the coefficient $\Phi$ is absent at NS order and thus it can be seen as an extension of a Burnett-order transport coefficient.
Figure \ref{fig8} depicts the $\alpha$-dependence of the reduced coefficients $\kappa/\kappa_0$ and $\Phi/\Phi_0$.
Interestingly, both reduced coefficients are quite similar for the whole range $\alpha_c\leq \alpha\leq 1$, where $\alpha_c\simeq 0.046$, the relative difference being smaller than 10\%. As expected, these coefficients diverge when $\alpha\to\alpha_c$ as a consequence of the similar divergence of the fourth-order moments in the USF. An important non-Newtonian effect (not directly observed in Fig.\ \ref{fig8}), is that, as $\alpha$ decreases, the magnitude of  the streamwise component $q_x$ grows more rapidly than that of the crosswise component $q_y$, so that the former becomes larger than the latter for  $\alpha \lesssim 0.9$, what represents a strong far-from-equilibrium effect. Finally, comparison between the generalized thermal conductivity $\kappa$ and the NS coefficient $\kappa_\NS$ shows that, in contrast to the cases of $\eta$ and $\eta_\NS$, both coefficients behave in a qualitatively similar way. In the range $0.475\leq\alpha<1$, $\kappa>\kappa_\NS$ (the relative difference being smaller than 20\%), while $\kappa<\kappa_\NS$ for $\alpha<0.475$.

\vspace*{0.5cm}
\setcounter{equation}{0}

\section{Small spatial perturbations around the USF\label{sec7}}
The LTu flow described in the preceding section can be seen as the USF perturbed by the existence of a thermal gradient parallel to the velocity gradient ($y$ axis) under the constraints of uniform pressure and heat flux. However, the perturbation is not small in the sense that the thermal gradient (as measured by the Knudsen number $\et$) is arbitrarily large.
In this section, we will adopt a complementary approach. On the one hand, the perturbations from USF will be assumed to be \textit{small}, but, on the other hand, they will affect \textit{all} the hydrodynamic fields.

Let us assume that the USF is disturbed by {small} spatial perturbations. The response of the system to these
perturbations gives rise to additional contributions to the momentum and heat fluxes, which can be characterized
by generalized transport coefficients \cite{G07}. Since the unperturbed system is strongly sheared, these generalized transport
coefficients are highly nonlinear functions of the shear rate. The goal here is to determine the shear-rate
dependence of these coefficients for IMM.

To analyze this problem, one has to start from the Boltzmann equation (\ref{1a}) with a general time and space
dependence. First, it is convenient to keep using the relative velocity ${\bf V}={\bf v}-{\bf u}_0$, where ${\bf
u}_0=ay\widehat{\mathbf{x}}$ is the flow velocity of the {\em undisturbed} USF state.  On the other hand, in the {\em disturbed}
state the true flow velocity ${\bf u}$ is in general different from ${\bf u}_0$, i.e., ${\bf u}={\bf u}_0+\delta {\bf
u}$, $\delta {\bf u}$ being a small perturbation to ${\bf u}_0$. As a consequence, the true peculiar velocity is
now  ${\bf W}\equiv {\bf v}-{\bf u}={\bf V}-\delta{\bf u}$. In the Lagrangian frame moving with velocity ${\bf
u}_0$, the Boltzmann equation \eqref{1a} reads
\begin{equation}
\label{3.1b}
\frac{\partial}{\partial t}f-aV_y\frac{\partial}{\partial V_x}f+\left({\bf V}+{\bf u}_0\right) \cdot
\nabla f=J[{\bf V}|f,f],
\end{equation}
where the gradient $\nabla f$ in the last term of the left-hand side must be taken  at constant ${\bf V}$.

The goal is to find a normal solution to Eq.\ \eqref{3.1b} that slightly deviates from the USF. For this
reason, let us assume that the spatial gradients of the hydrodynamic fields
\begin{equation}
\label{3.1.1}
A({\bf r},t)\equiv \{n({\bf r},t), T({\bf r}, t),
\delta {\bf u}({\bf r},t)\}
\end{equation}
are small. Under these conditions, it is appropriate to solve Eq.\ (\ref{3.1b})  by means of
a generalization of the conventional Chapman--Enskog method \cite{CC70}, where the velocity distribution
function is expanded about a {\em local} shear flow reference state. This type of Chapman--Enskog-like expansion has been
considered in the case of elastic gases to get the set of shear-rate dependent transport coefficients
 in a thermostatted shear flow problem \cite{GS03,LD97} and  has also been  considered
 in the context of inelastic gases \cite{G06,G07,G07b,L06}.

As said in section \ref{sec4}, the Chapman--Enskog method assumes the existence of a {\em normal} solution in which all space and time
dependence of the distribution function occurs through a functional dependence on the fields $A({\bf r},t)$,
i.e.,
\begin{equation}
\label{3.2} f= f[{\bf V}|A].
\end{equation}
 This functional dependence can be made local by an
expansion of the distribution function in powers of the
hydrodynamic gradients:
\begin{equation}
\label{3.3} f({\bf V}) =f^{(0)}({\bf V}|A)+ \epsilon f^{(1)}({\bf V}|A)+\cdots,
\end{equation}
where, as in Eq.\ \eqref{c1}, $\epsilon$ is a bookkeeping  parameter that can be set equal to 1 at the end of the calculations. The reference zeroth-order distribution function corresponds to the \textit{unsteady} USF distribution function but taking
into account the local dependence of the density and temperature and the change ${\bf V}\rightarrow {\bf W}={\bf
V}-\delta{\bf u}({\bf r}, t)$. It is important to note that, as seen in section \ref{sec5a}, in the stationary USF the temperature is fixed by the shear rate and the coefficient of restitution [cf.\ Eq.\ \eqref{5.7}]. Therefore, in order to have $T$ as an independent field, one needs to solve the time-dependent USF problem, as discussed in section \ref{sec5b} As a consequence, the associated solution $f^{(0)}$, in dimensionless form, is a function of $\alpha$ and $a^*\equiv a/\nu_0$ separately. Apart from this difficulty, a new feature of the Chapman--Enskog-like expansion \eqref{3.3} (in contrast to the
conventional one) is that the successive approximations $f^{(k)}$ are of order $k$ in the gradients of $n$, $T$,
and $\delta {\bf u}$, but retain all the orders in the reduced shear rate $a^*$ \cite{G07}.

The expansion (\ref{3.3}) yields the corresponding expansions for
the fluxes:
\begin{equation}
\label{3.4} {\sf P}={\sf P}^{(0)}+\epsilon {\sf P}^{(1)}+\cdots, \quad {\bf
q}=\epsilon {\bf q}^{(1)}+\cdots,
\end{equation}
where $P_{ij}^{(0)}=p P_{ij}^*(\al,a^*)$ is the pressure tensor in the \textit{unsteady} USF and we have taken into account that ${\bf q}^{(0)}=0$.
A careful application of the Chapman--Enskog-like expansion to first order gives the following forms for the generalized constitutive equations \cite{G07}:
\begin{equation}
\label{3.22} P_{ij}^{(1)}=-\eta_{ijk\ell} \nabla_\ell \delta
u_k,
\end{equation}
\begin{equation}
\label{3.23} q_i^{(1)}=-\kappa_{ij}\nabla_j T- \mu_{ij}\nabla_j n.
\end{equation}
In general, the set of {\em generalized} transport coefficients
$\eta_{ijk\ell}$, $\kappa_{ij}$, and $\mu_{ij}$ are nonlinear
functions of the coefficient of restitution $\alpha$ and the
reduced shear rate $a^*$. The anisotropy induced in the system by
the presence of shear flow gives rise to new transport
coefficients, reflecting broken symmetry. The momentum flux is
expressed in terms of a viscosity tensor $\eta_{ijk\ell}(a^*,
\alpha)$ of rank 4 which is symmetric and traceless in $ij$ due to
the properties of the pressure tensor $P_{ij}^{(1)}$. The heat
flux is expressed in terms of a thermal conductivity tensor
$\kappa_{ij}(a^*, \alpha)$ and a new tensor $\mu_{ij}(a^*,
\alpha)$. Of course, for $a^*=0$ and $\alpha=1$, the usual
NS constitutive equations for ordinary gases are
recovered and the transport coefficients become
\begin{equation}
\label{3.27} \eta_{ijk\ell}\rightarrow
\eta_0\left(\delta_{ik}\delta_{j\ell}+\delta_{jk}\delta_{i\ell}-
\frac{2}{d}\delta_{ij}\delta_{k\ell}\right),\quad
\kappa_{ij}\rightarrow \kappa_0 \delta_{ij}, \quad
\mu_{ij}\rightarrow 0.
\end{equation}

The elements of the tensor $\eta_{ijk\ell}$ obey a set of coupled linear first-order differential equations in terms of $a^*$, $P_{ij}^*$, and $\partial_{a^*}P_{ij}$.  Those differential equations become algebraic equations when one specializes to the \textit{steady-state} condition \eqref{5.2}. In that case one gets
\begin{eqnarray}
\label{4.5}
a^*\left(\delta_{ix}\eta_{jyk\ell}^*+
\delta_{jx}\eta_{iyk\ell}^*-\delta_{ky}\eta_{ijx\ell}^*\right)
+\nu_{0|2}^*\eta_{ijk\ell}^*
&=&\delta_{k\ell}a^*\partial_{a^*}P_{ij}^{*}+
\delta_{ik}P_{j\ell}^{*}+\delta_{jk}P_{i\ell}^{*}\nonumber\\
& &
-\frac{2}{d}\left(P_{k\ell}^{*}-a^*\eta_{xyk\ell}^*\right)\left(1-\frac{a^*}{2}\partial_{a^*}\right)
P_{ij}^{*},\nn
\end{eqnarray}
where $\eta_{ijk\ell}^*\equiv \eta_{ijk\ell}/\eta_0$. In Eq.\ \eqref{4.5}, $a^*$, $P_{ij}^*$, and $\partial_{a^*} P_{ij}^*$ are functions of $\al$ given by Eqs.\ \eqref{5.6}, \eqref{5.7}--\eqref{5.10}, \eqref{5.22}, and \eqref{5.23}. As a simple test, note that in the elastic limit ($a^*=0$, $\nuzt^*=1$, $P_{ij}^*=\delta_{ij}$), Eq.\ \eqref{4.5} becomes Eq.\ \eqref{3.27}.
Also, it must be noted that, on physical grounds, the elements of the form $\eta_{ijxy}$ are directly related to the \textit{unperturbed}  transport coefficients $\eta$ and $\Psi$ defined by Eqs.\ \eqref{5.11} and \eqref{5.12}. The rationale is that the particular case of a perturbation in the velocity field of the form $ \delta \mathbf{u}=(\delta a) y\widehat{\mathbf{x}}$ is totally equivalent to an unperturbed USF state with a shear rate $a+\delta a$. As a consequence, one has
\beqa
P_{xy}^*(a^*+\delta a^*)&=&-\eta^*(a^*+\delta a^*)\left(a^*+\delta a^*\right)\nn
&=&-\eta^*(a^*)a^*-\delta a^*\left(1+a^*\partial_a^*\right)\eta^*(a^*)+\cdots,
\label{7.1}
\eeqa
\beqa
P_{yy}^*(a^*+\delta a^*)=P_{zz}^*(a^*+\delta a^*)&=&1-\frac{1}{d}\Psi^*(a^*+\delta a^*)\left(a^*+\delta a^*\right)^2\nn
&=&1-\frac{1}{d}\Psi^*(a^*){a^*}^2-\delta a^*\frac{2a^*}{d}\left(1+\frac{a^*}{2}\partial_a^*\right)\Psi^*(a^*)+\cdots,\nn
\label{7.2}
\eeqa
\beq
P_{xx}^*(a^*+\delta a^*)=d-(d-1)P_{yy}^*(a^*+\delta a^*).
\label{7.3}
\eeq
This implies that
\beq
\eta_{xyxy}^*(a^*)=\left(1+a^*\partial_a^*\right)\eta^*(a^*) ,
\label{7.4}
\eeq
\beq
\eta_{yyxy}^*(a^*)=\eta_{zzxy}^*(a^*)=\frac{2a^*}{d}\left(1+\frac{a^*}{2}\partial_a^*\right)\Psi^*(a^*),
\label{7.5}
\eeq
\beq
\eta_{xxxy}^*(a^*)=-(d-1)\eta_{yyxy}^*(a^*).
\label{7.6}
\eeq
Here, $\eta^*\equiv \eta/\eta_0$, $\delta a^*\equiv \delta a/\nu_0$, and $\Psi^*\equiv \Psi/(\eta_0^2/p)$. Equations \eqref{7.4}--\eqref{7.6} hold both for the steady and unsteady USF. It can be checked that they are consistent with Eq.\ \eqref{4.5} in the steady state, in which case $\partial_{a^*}\eta^*$ and $\partial_{a^*}\Psi^*$ are obtained from Eqs.\ \eqref{5.22} and \eqref{5.23}. The elements of the steady-state shear viscosity tensor $\eta^*_{ijk\ell}$ with $k\neq x$ and $\ell\neq y$ must be obtained by solving the set of algebraic equations \eqref{4.5}.

It turns out that there are two classes of terms. Class I is made of those coefficients $\eta^*_{ijk\ell}$ with  $(k,\ell)=(xx)$, $(xy)$, $(yx)$, $(yy)$, and $(zz)$. The complementary class II includes the coefficients with  $(k,\ell)=(xz)$, $(yz)$, $(zx)$, and $(zy)$. Of course, class II (as well as the elements  $\eta^*_{ijzz}$ of class I) are meaningless if $d=2$. The coefficients of the form $\eta^*_{xzk\ell}$ and $\eta^*_{yzk\ell}$ vanish in class I, while those of the form $\eta^*_{xxk\ell}$, $\eta^*_{xyk\ell}$, and $\eta^*_{yyk\ell}$ vanish in class II. The remaining elements in class II are
\beq
\eta_{xzxz}^*=\eta^*,\quad \eta_{yzxz}^*=0,
\label{7.7}
\eeq
\beq
\eta_{xzyz}^*=0,\quad \eta_{yzyz}^*=\eta^*,
\label{7.8}
\eeq
\beq
\eta_{xzzx}^*=\eta^*\left[\frac{P_{xx}^*}{P_{yy}^*}+\left(\frac{P_{xy}^*}{P_{yy}^*}\right)^2\right],
\quad \eta_{yzzx}^*=\eta^*\frac{P_{xy}^*}{P_{yy}^*},
\label{7.9}
\eeq
\beq
\eta_{xzzy}^*=2\eta^*\frac{P_{xy}^*}{P_{yy}^*},\quad \eta_{yzzy}^*=\eta^*.
\label{7.10}
\eeq
Some of the above results might have been anticipated from simple arguments, as shown on p.\ 138 of Ref.\ \cite{GS03}.

The expressions for the non-zero elements of class I include the derivatives $\partial_{a^*}\eta^*$ and $\partial_{a^*}\Psi^*$. Those expressions are much more involved than those of class II and so they will not be explicitly given here, except for the cases of Eqs.\ \eqref{7.4}--\eqref{7.6}. As Eq.\ \eqref{7.6} shows, the combination $\eta_{xxk\ell}^*+(d-1)\eta_{yyk\ell}^*$ vanishes for $(k,\ell)=(xy)$. It also does for $(k,\ell)=(xx)$, while for the other cases of class I one simply has
\beq
\eta_{xxyx}^*+(d-1)\eta_{yyyx}^*=2(d-2)\eta^*\frac{P_{xy}^*}{P_{yy}^*},
\label{7.11}
\eeq
\beq
\eta_{xxyy}^*+(d-1)\eta_{yyyy}^*=2(d-2)\eta^*,
\label{7.12}
\eeq
\beq
\eta_{xxzz}^*+(d-1)\eta_{yyzz}^*=-2\eta^*.
\label{7.13}
\eeq

\begin{figure}[htbp]
\centerline{\includegraphics[width= .5\columnwidth]{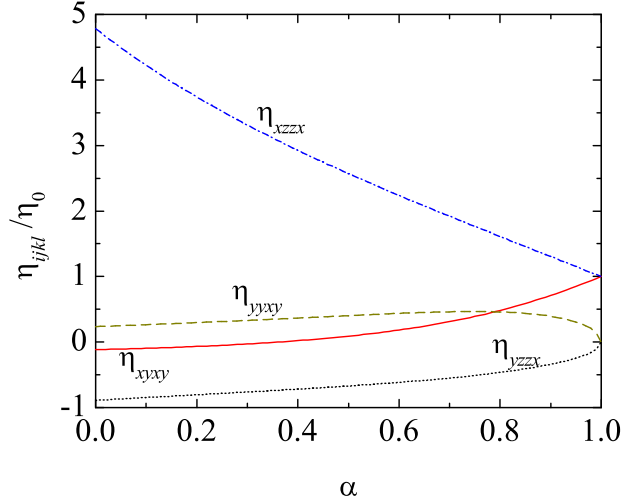}}
\caption{Plot of the reduced generalized transport coefficients $\eta_{xyxy}/\eta_0$ (solid line), $\eta_{yyxy}/\eta_0$ (dashed line), $\eta_{xzzx}/\eta_0$ (dash-dotted line), and $\eta_{yzzx}/\eta_0$ (dotted line) in the steady-state USF for $d=3$.\label{fig9}}
\end{figure}

Figure \ref{fig9} shows the steady-state values of  two elements of class I ($\eta_{xyxy}^*$ and $\eta_{yyxy}^*$) and two elements of class II ($\eta_{xzzx}^*$ and $\eta_{yzzx}^*$) in the three-dimensional case. We recall that the first two coefficients measure the deviations of $P_{xy}$ and $P_{yy}$ with respect to their unperturbed USF values due to a perturbation of the form $\partial\delta {u}_x/\partial y$. Analogously, the two coefficients $\eta_{xzzx}^*$ and $\eta_{yzzx}^*$ measure the presence of non-zero values of $P_{xz}$ and $P_{yz}$, respectively, due to a perturbation of the form $\partial\delta {u}_z/\partial x$.
We observe that, at a given value of $\alpha$, the largest influence occurs on $P_{xz}$. It is also interesting to note that $\eta_{xyxy}^*$ becomes negative at strong values of dissipation, while $\eta_{yzzx}^*$ is always negative.

The evaluation of the heat flux coefficients $\kappa_{ij}$ and $\mu_{ij}$ is more involved than that of the shear viscosity tensor $\eta_{ijk\ell}$. In the general unsteady case, $\kappa_{ij}$ and $\mu_{ij}$ obey coupled linear first-order differential equations where, in addition to $a^*$, $P_{ij}^*$, and $\partial_{a^*}P_{ij}$, the fourth-order moments of the USF and their first derivatives with respect to $a^*$ are also involved \cite{G07}. For steady-state  conditions, the set of equations becomes algebraic.
For further analysis, let us rewrite Eq.\ \eqref{3.23} as
\begin{equation}
\label{3.23b}
q_i^{(1)}=-\widetilde{\kappa}_{ij}\nabla_j T- \widetilde{\mu}_{ij}\nabla_j n+\gamma_{ij}\nabla a^*,
\end{equation}
so that
\beq
\kappa_{ij}=\widetilde{\kappa}_{ij}+\frac{a^*}{2T}\gamma_{ij},\quad \mu_{ij}=\widetilde{\mu}_{ij}+\frac{a^*}{n}\gamma_{ij}.
\label{7.15}
\eeq
In Eq.\ \eqref{3.23b} we have disentangled the contributions to the heat flux directly associated with the temperature and density gradients from those due to the local spatial dependence of $a^*\propto n^{-1}T^{-1/2}$. Whereas the coefficients $\widetilde{\kappa}_{ij}$ and $\widetilde{\mu}_{ij}$ are given in terms of the second- and fourth-order moments of USF, but not of their derivatives with respect to $a^*$, the coefficients $\gamma_{ij}$ are linear functions of those derivatives.
The equations for $\kappa_{ij}$ and $\mu_{ij}$ are not reproduced here and the interested reader is referred to Ref.\ \cite{G07}.

It is illuminating to connect the elements $\widetilde{\kappa}_{ij}$ and $\widetilde{\mu}_{ij}$ in the steady state with the LTu transport coefficients $\kappa$ and $\Phi$ defined by Eq.\ \eqref{Z14}. The key point is the realization that the LTu (see Sec.\ \ref{sec6}) can be interpreted as a special perturbation of the USF such that (a) the only non-zero temperature and density gradients are $\partial_y T$ and $\partial_y n$, (b) those gradients are not independent but are related by the constant-pressure condition $\partial_y n=-(n/T)\partial_y T$, and (c) the reduced shear rate $a^*$ is constant. Although in the LTu the strength of the ``perturbation'' $\partial_y T$ is arbitrary, we have seen that the heat flux is linear in the thermal gradient, so that the effective coefficients defined by Eq.\ \eqref{Z14} must be related to those defined by \eqref{3.23b} as
\beq
\kappa=\widetilde{\kappa}_{yy}-\frac{n}{T}\widetilde{\mu}_{yy},\quad -\Phi a=\widetilde{\kappa}_{xy}-\frac{n}{T}\widetilde{\mu}_{xy}.
\label{7.14}
\eeq
We have checked that the above relations are indeed satisfied, what shows the consistency of our results.

\vspace*{0.5cm}
\setcounter{equation}{0}

\section{Concluding remarks\label{sec8}}

Exact solutions in nonequilibrium statistical mechanics are scarce. In order to overcome this limitation,  two possible alternatives can be envisaged from a theoretical point of view. On the one hand, one can consider a realistic and detailed description but make use of approximate (and sometimes uncontrolled) techniques to get quantitative predictions. On the other hand, one can introduce an idealized mathematical model (which otherwise captures the most relevant physical properties of the underlying system) and solve it by analytical and exact methods.
Here we have adopted the second strategy.

In this review we have offered an overview of some recent \textit{exact} results obtained in the context of the Boltzmann equation for a granular gas modeled as inelastic Maxwell particles.
Although most of the results reviewed in this paper have been reported previously, some other ones are original and presented here for the first time.

We have focused on the hydrodynamic properties of the system, where here the term `hydrodynamics' has been employed in a wide sense encompassing both Newtonian and non-Newtonian behavior.
More specifically, the Navier--Stokes (NS) transport coefficients $\eta_\NS$, $\kappa_\NS$, $\mu_\NS$, and $D_\NS$ have been obtained from the Chapman--Enskog method in Sec.\ \ref{sec4} As a common feature, it is observed that the collisional inelasticity produces an increase of all the NS transport coefficients, especially those related to the heat flux. In fact, the latter coefficients diverge (in two and three dimensions) for sufficiently small values of the coefficient of restitution $\alpha$. This is a consequence of the algebraic high-velocity tail of the distribution function in the homogeneous cooling state.

As an example of non-Newtonian hydrodynamics, the study of the paradigmatic uniform shear flow (USF) has been addressed in Sec.\ \ref{sec5} The analysis includes both the steady and (transient) unsteady states. While the former  has been extensively studied in the literature, the hydrodynamic transient toward the steady state has received much less attention. We have studied the rheological properties (generalized shear viscosity $\eta$ and viscometric function $\Psi$), thus assessing the influence of inelasticity on momentum transport. Additionally, the $\alpha$-dependence of the fourth-order velocity cumulant $a_2$ in the steady state has been shown. Again, an algebraic high-velocity tail of the USF distribution function gives rise to a divergence of $a_2$ for quite small values of $\alpha$ ($\alpha\lesssim 0.046$ for $d=3$).

Next, a more complex state has been analyzed in Sec.\ \ref{sec6} This state (referred to as LTu) is actually a class of planar Couette flows characterized by a uniform heat flux, as a consequence of an exact balance between viscous heating and collisional cooling contributions. In contrast to the USF (where only momentum flux is present) and to the Fourier flow for an ordinary gas (where only heat flux is present), both momentum and heat fluxes coexist in this class of states. It turns out that the rheological properties coincide with those of the USF. The most interesting  result is that the heat flux
is exactly proportional to the thermal gradient with coefficients $\kappa$ and $\Phi$ that are highly nonlinear functions of $\alpha$.

Finally, the problem of arbitrary (but small) spatial perturbations to the USF has been considered in Sec.\ \ref{sec7} Taking the USF  as a reference state and carrying out a Chapman--Enskog-like expansion about it, a generalized shear viscosity tensor $\eta_{ijk\ell}$ and  generalized heat-flux tensors $\kappa_{ij}$ and $\mu_{ij}$ are determined. The $\alpha$-dependence of $\eta_{xyxy}$, $\eta_{yyxy}$, $\eta_{xzzx}$, and $\eta_{yzzx}$ has been explicitly given.

It is worth emphasizing that all the results reviewed in this paper are \textit{exact} in the context of the Boltzmann equation for IMM, regardless of the degree of dissipation. Moreover, all the results are explicit, with the exception of those displayed in Fig.\ \ref{fig7b}, which are obtained from a numerical solution of the set of coupled differential equations  \eqref{5.17} and \eqref{5.18}. This contrasts with the case of inelastic hard spheres, which requires the use of approximations and/or numerical methods. The price to be paid is that, in general, the quantitative predictions obtained from IMM  significantly enhance the influence of dissipation on the dynamical properties of a granular gas.  However, the IMM is very useful to unveil in a clean way the role played by inelasticity in granular flows, especially in situations, such as highly non-Newtonian states, where simple intuition is not enough.

\vspace*{0.5cm}

\section*{Acknowledgements}
This work  has been supported by the Ministerio de  Ciencia e Innovaci\'on (Spain) through Grant No.\ FIS2010-16587
(partially financed by FEDER funds) and by the Junta de Extremadura (Spain) through Grant No.\ GR10158.


\end{document}